\documentclass[11pt]{article}
\usepackage[utf8]{inputenc}
\usepackage[numbers,sort&compress]{natbib} 
\usepackage[a4paper,left=2.54cm,right=2.54cm,top=3.17cm,bottom=3.17cm]{geometry} 
\usepackage{setspace} 
\usepackage{graphicx} 
\usepackage{float} 
\usepackage{subfigure} 
\usepackage{caption} 
\usepackage{hyperref} 
\usepackage{amsmath} 
\usepackage{amsfonts} 
\usepackage{xfrac} 
\usepackage{amssymb}
\usepackage{upgreek}
\usepackage{makecell}
\usepackage{tabularx}
\usepackage{booktabs}
\usepackage{array}
\usepackage{authblk}

\usepackage{multirow}
\usepackage{multicol}

\title{Fiber-optic multimode interference sensing: comprehensive characterization and its potential for strain-insensitive temperature sensing 
}
\author{Kun~Wang$^{1,}$*, Yosuke~Mizuno$^2$, Xingchen~Dong$^1$,  Wolfgang~Kurz$^1$, Maximilian~Fink$^1$, Heeyoung Lee$^3$, Martin~Jakobi$^1$, and~Alexander~W.~Koch$^1$}
\affil{\small{1. Technical University of Munich, Germany, Department of Electrical and Computer Engineering, Institute for Measurement Systems and Sensor Technology, Munich 80333, Germany\\
2. Yokohama National University, Faculty of Engineering, Yokohama 240-8501, Japan\\
3. College of Engineering, Shibaura Institute of Technology, Tokyo 135-8548, Japan}}
\date{}

\begin{document} 
\maketitle

\begin{abstract}
\noindent A strain-insensitive temperature sensor based on multimode interference using standard multimode fibers (MMFs) is proposed according to the comprehensive study of the characteristics of the MMFs. The temperature and strain dependences on the core diameter, numerical aperture (NA), and the length of the MMF section in the single-mode--multimode--single-mode (SMS) fiber structure are investigated experimentally. The results indicate that the larger core diameter of the MMF leads to higher temperature sensitivity but lower strain sensitivity (absolute values); the higher NA does not influence the temperature sensitivity but results in higher absolute value of strain sensitivity; the longer MMF section brings lower temperature sensitivity but does not have an impact on strain sensitivity. These findings also contribute to the theoretical analysis of the length dependence in the SMS fiber sensors. Besides, the results of the characterization study show that the strain sensitivity is relatively low, which brings a possibility to develop a strain-insensitive temperature sensor. The proposed sensor is used for temperature sensing while the strain is constantly applied from 0 to 1100 $\upmu\upvarepsilon$ with steps of 100 $\upmu\upvarepsilon$. The measured results are consistent with the comprehensive study. The mean temperature sensitivity is 6.14~pm/$^{\circ}$C with a standard deviation of 0.39~pm/$^{\circ}$C, which proves that the proposed temperature sensor exhibits good stability and is insensitive to strain. We expect that these results will provide a profound guideline to fiber sensors based on multimode interference.	
\end{abstract}

\section{Introduction}\label{Section 1}
Optical fiber sensors have been a promising sensing device for strain and temperature measurement for decades due to their intrinsic characteristics, such as compact size, lightweight, fast response, remote sensing capabilities, and insensitivity to ambient electromagnetic fields~\cite{LEUNG2007688, 9268078, analytical_chemistry, Chen:10, me}. 
Strain and temperature sensing using optical fibers have received significant attention in research, and various fiber sensors have been developed by exploiting fiber Bragg gratings (FBGs)~\cite{Guan, OliveiraFBG}, long-period gratings (LPGs)~\cite{Wang:06, ZhaoLPG}, Raman scattering~\cite{Alahbabi:05}, Brillouin scattering~\cite{Minakawa2017, Fang:19}, and surface plasmon resonance~\cite{Wang:18}. One of the simple and low-cost implementations based on the multimode interference (MMI) effect is the so-called single-mode--multimode--single-mode (SMS) fiber sensor. The SMS fiber sensor is composed of a short segment of a multimode fiber (MMF) sandwiched between two single-mode fibers (SMFs), which was firstly reported by Mehta et al. in 2003~\cite{Mehta}. Since then, various optical fiber sensors based on MMI have been investigated, as summarized in~\cite{Frazao:11, me}. 

Based on the SMS fiber structure, the sensitivity dependences on the fiber characteristics, such as length and core diameter, are investigated. Qiang Wu et al.~\cite{Wu:11} reported the influence of the MMF core diameter and lengths on the refractive index (RI) sensitivity of an SMS fiber sensor using uncladded MMF. The results showed that the smaller core diameter of MMF leads to a higher sensitivity, but the length does not influence the sensitivity. 
Later, Yang Li et al.~\cite{li2014multimode} reported a multimode interference refractive index sensor using a coreless fiber. The results showed that the length of the coreless fiber does not influence the RI sensitivity, but the smaller core diameter tends to have a higher RI sensitivity. Similarly, Yaofei Chen et al.~\cite{chen2014} reported the same conclusion by demonstrating an all-fiber refractometer based on SMS structure implementing a no-core fiber (NCF). However, the temperature and strain characteristics of the SMS structure comprising standard MMFs have not been systematically investigated yet.   

In this work, we perform a comprehensive characterization of the standard MMF-based SMS structure for temperature and strain sensing. Then, on the basis of the results, we show its potential for strain-insensitive temperature sensing. The characterization for temperature and strain sensing includes the following: 
\begin{enumerate}
	\item The dependence on the core diameter of the MMF
	\item The dependence on the NA of the MMF
	\item The dependence on the length of the MMF
\end{enumerate}

\noindent This detailed study also shows that the strain sensitivity is relatively small and thus has the potential to develop strain-insensitive temperature sensors with standard MMFs instead of specialty fibers, such as small-core photosensitive fiber~\cite{CAO201424} and seven-core fiber~\cite{LIU2019172}.
According to this study, the strain-insensitive temperature fiber sensor based on MMI implementing standard MMF is verified experimentally.

\section{Principle}
The SMS structure consists of a short section of an MMF whose ends are connected to the SMFs. The injected light is guided from the input SMF into the MMF and propagates along with the MMF. At the first SMF/MMF boundary, the spot-size difference between the fundamental modes in the SMF and MMF excites a few lower modes in the MMF, which propagate with different propagation constants~\cite{Mohammed:06}. At the second SMF/MMF boundary, the net field coupled to the output SMF is determined by the relative phase diﬀerences among the many modes guided in the MMF.~\cite{YosukeMIZUNO2018} Based on the assumption that the MMF and SMFs are axially aligned, the modes excited in the MMF are axially symmetric, and according to the detailed calculation~\cite{kumar2003, Tripathi2009}, the sensing principle can be summarized as 
\begin{equation}
	P_{out} = |\,A_0^2 + A_1^2 \, e^{i(\beta_0-\beta_1)L} + A_2^2 \, e^{i(\beta_0-\beta_2)L} + \dots|^2 \,,
	\label{equ.1}
\end{equation} 
where $P_{out}$ is the power in the output SMF, $A_i$ is defined as the field amplitude of the $i$-th mode at the first SMF/MMF boundary, $\beta_i$ is the propagation constant of the $i$-th mode, and $L$ is the length of the MMF. The propagation constant of the $i$-th symmetric mode can be obtained as 
\begin{equation}
	\beta_i = k_0 \, n_{co} \left[ 1-\frac{2\,(2\,i-1)\,\alpha_M}{k_0^2\,n_{co}^2} \right]^{1/2}
	\label{equ.2}
\end{equation}

\noindent with

\begin{equation}
	\alpha_M = \frac{k_0 \,(NA)}{a_M},
	\label{equ.3}
\end{equation}

\noindent where $i = 0,1,2,\dots$, $k_0$ is the free-space wave number, $n_{co}$ represents the refractive index of the MMF core, and $a_M$ is defined as the core radius.
Equations~(\ref{equ.1}), (\ref{equ.2}), and (\ref{equ.3}) show that the optical power in the output fiber is affected by changes in the propagation constants, i.e., the NA, core diameter, and the length of the MMF, caused by the physical quantities temperature and strain in this work. Therefore, these changes can be quantitatively evaluated by measuring the shift in either power or spectral location of peaks (or dips). For simplicity, this theory supposes the interference of the light only includes two modes (fundamental and first-order high modes~\cite{kumar2003}) propagating along with the MMF. However, in the experiment, not only these two modes but also many other modes need to be simultaneously considered~\cite{Tomohito2017}. In this work, besides the theoretical analysis, a comprehensive study including the core diameter, NA, and length of the standard MMF is performed. It will give a valuable guideline for future fabrication of SMS fiber sensors.

\section{Experimental Results and Discussion}
\subsection{Experiment}
\begin{figure}[]
	\centering
	\includegraphics[scale=0.6]{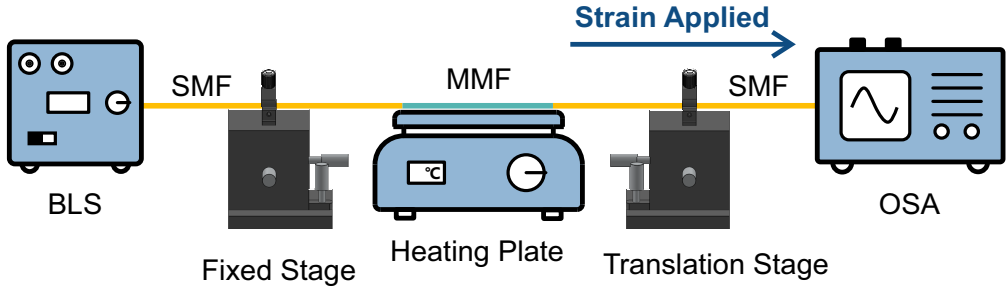}
	\caption{Schematic diagram of the experimental setup for temperature and strain measurement; BLS, broadband light source; SMF, single-mode fiber; MMF, multimode fiber; OSA, optical spectrum analyzer.}
	\label{fig:setup}
\end{figure}

Figure~\ref{fig:setup} shows the schematic diagram of the experimental setup for temperature and strain measurement. The MMF section is placed on a heating plate and clamped on two precision translation stages, which can apply varying axial tensile stress on the sample. The applied axial strain can be calculated by~\cite{Li:07}

\begin{equation}
	\upvarepsilon = \frac{\Delta L}{L}\,,
	\label{equ.4}
\end{equation} 

\noindent where $L$ is the initial length and $\Delta L$ is the additional change in length of the MMF section when longitudinal stress is introduced. Both ends of the MMF are connected to the SMFs. The other end of the input SMF is connected to a broadband light source (BLS, central wavelength: 1550 nm), while the other end of the output SMF is connected to an optical spectrum analyzer (OSA), which detects the changes in the light spectrum. 

The temperature sensitivity was first measured without strain applied during the experiment. Due to the different operating temperatures of the MMFs, the temperature range was chosen from 30 to 70$^{\circ}$C in steps of 5$^{\circ}$C. 
Then the strain sensitivity was measured at room temperature $\sim$20$^{\circ}$C. The strain range was 0 to 1750~$\upmu\upvarepsilon$ according to the available devices in steps of 250~$\upmu\upvarepsilon$.

\subsection{Core Diameter Dependence}
The commonly used standard MMF has a 0.22~NA silica core with three different core diameters: 50~$\upmu$m (FG050LGA, Thorlabs), 105~$\upmu$m (FG105LGA, Thorlabs), and 200~$\upmu$m (FG200LGA, Thorlabs). Therefore, these three MMFs are used for investigating the core diameter dependence with a length of 8 cm, which is determined based on the length dependence in the following subsection. 

\begin{figure}[]
	\centering
	\vspace{-0.35cm}
	\subfigtopskip=2pt
	\subfigbottomskip=2pt
	\subfigcapskip=-5pt
	\subfigure[50-$\upmu$m core]{
		\begin{minipage}[t]{0.32\columnwidth}
			\includegraphics[width=1\linewidth]{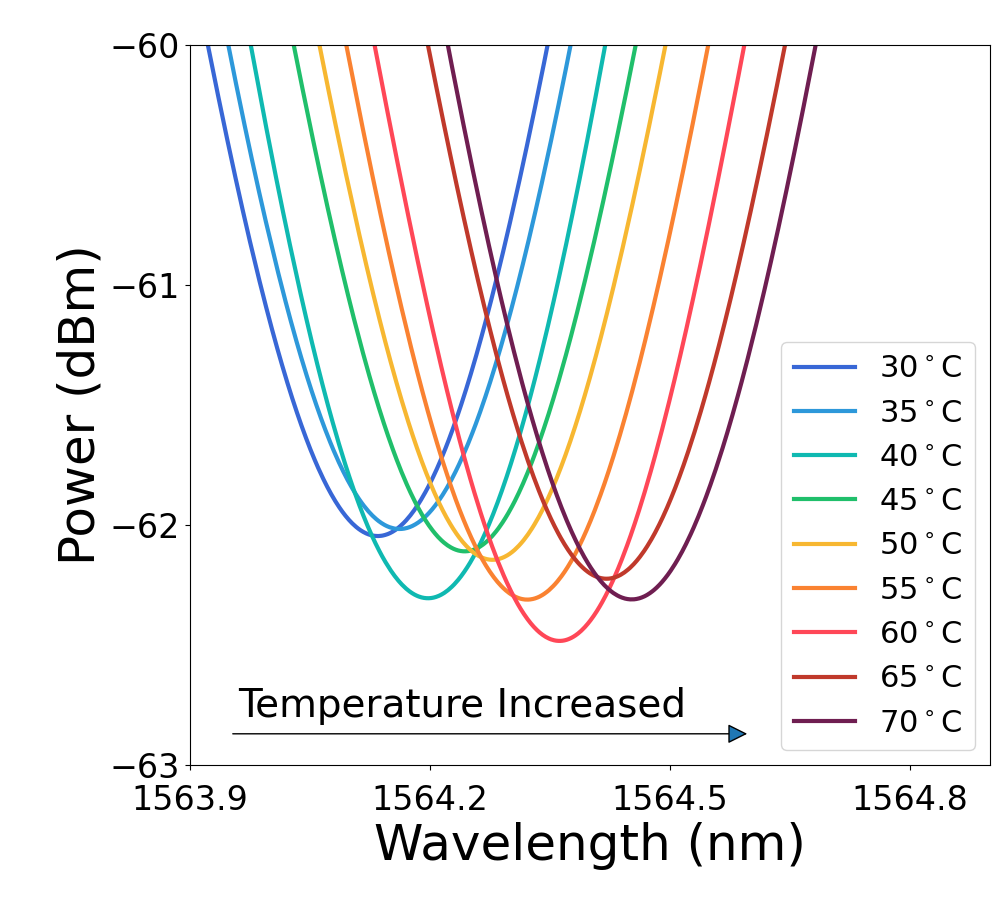}
			\label{fig.core50.wavelength_temp}
	\end{minipage}}
	\subfigure[105-$\upmu$m core]{
	\begin{minipage}[t]{0.32\columnwidth}
		\includegraphics[width=1\linewidth]{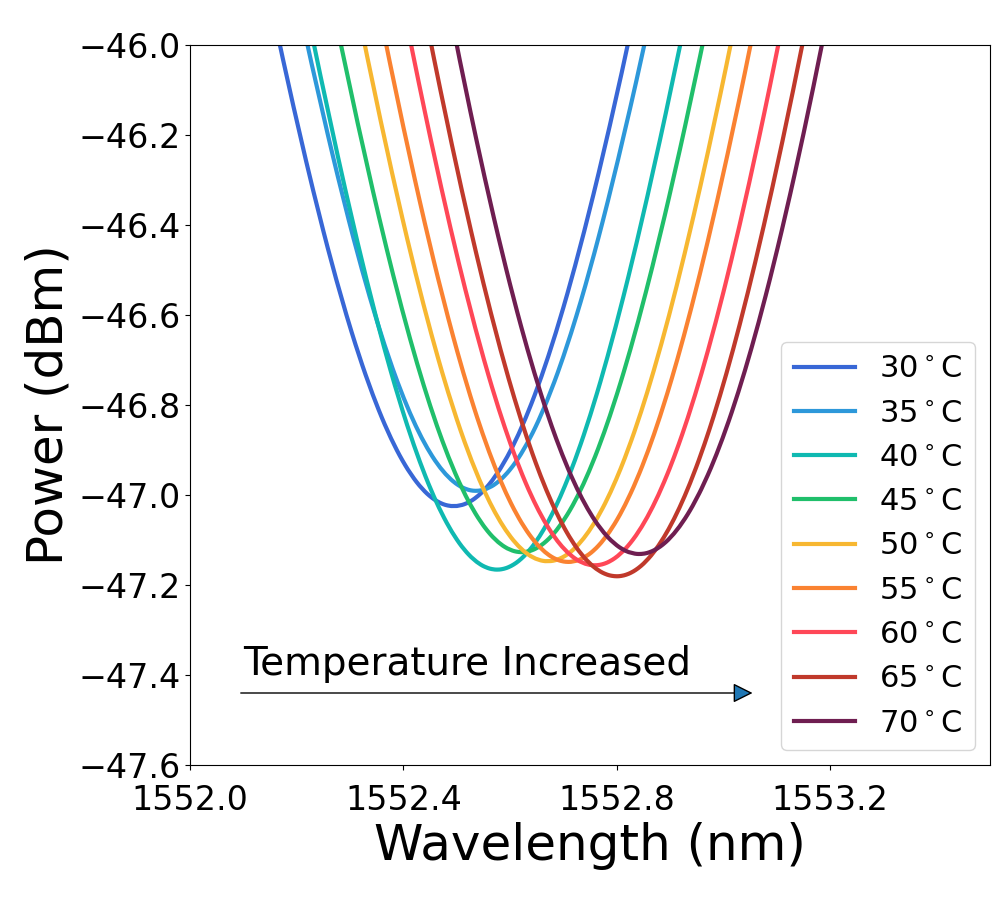}
		\label{fig.core105.wavelength_temp}
	\end{minipage}}
	\subfigure[200-$\upmu$m core]{
	\begin{minipage}[t]{0.32\columnwidth}
		\includegraphics[width=1\linewidth]{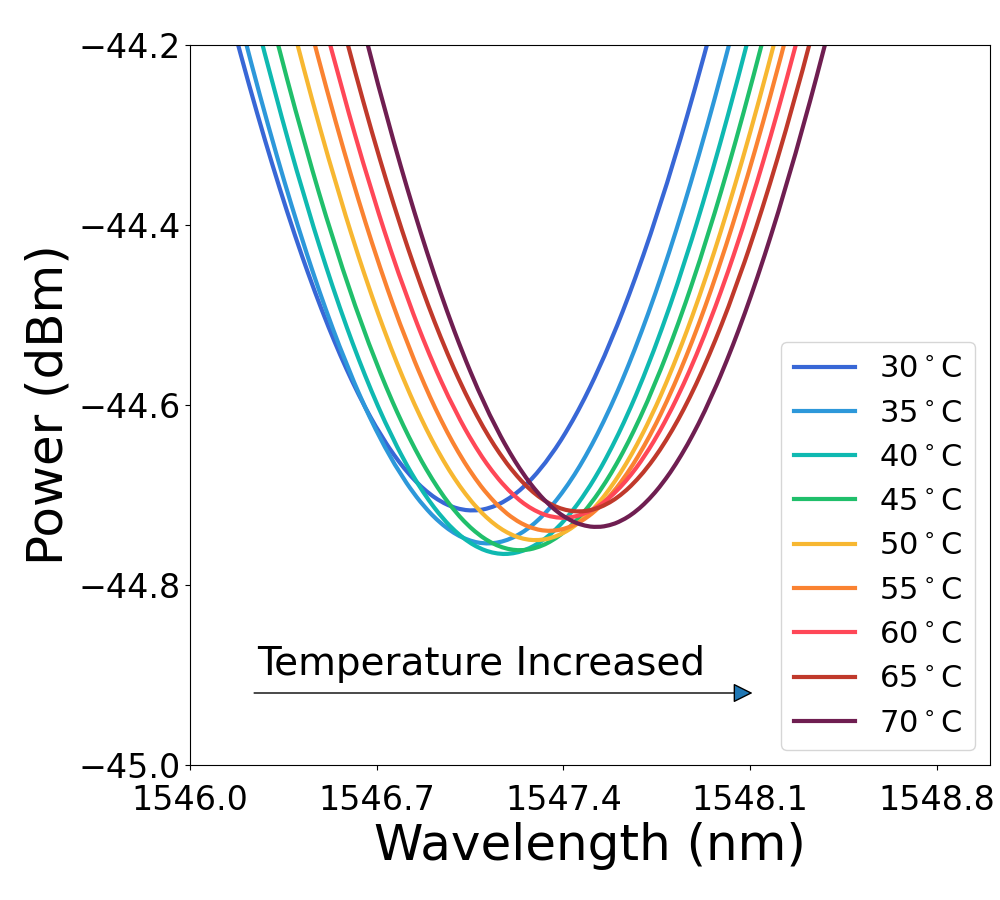}
		\label{fig.core200.wavelength_temp}
	\end{minipage}}
\\
	\subfigure[50-$\upmu$m core]{
		\begin{minipage}[t]{0.32\columnwidth}
			\includegraphics[width=1\linewidth]{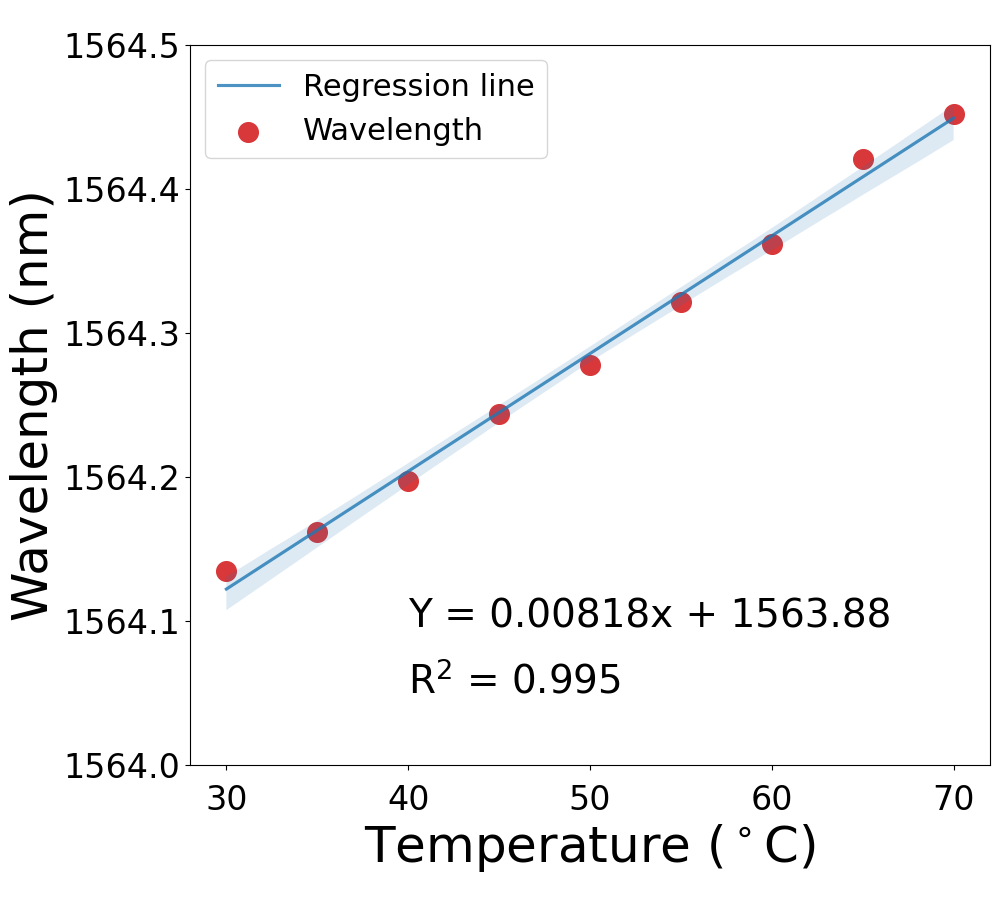}
			\label{fig.core50.fitting_temp}
	\end{minipage}}
	\subfigure[105-$\upmu$m core]{
		\begin{minipage}[t]{0.32\columnwidth}
			\includegraphics[width=1\linewidth]{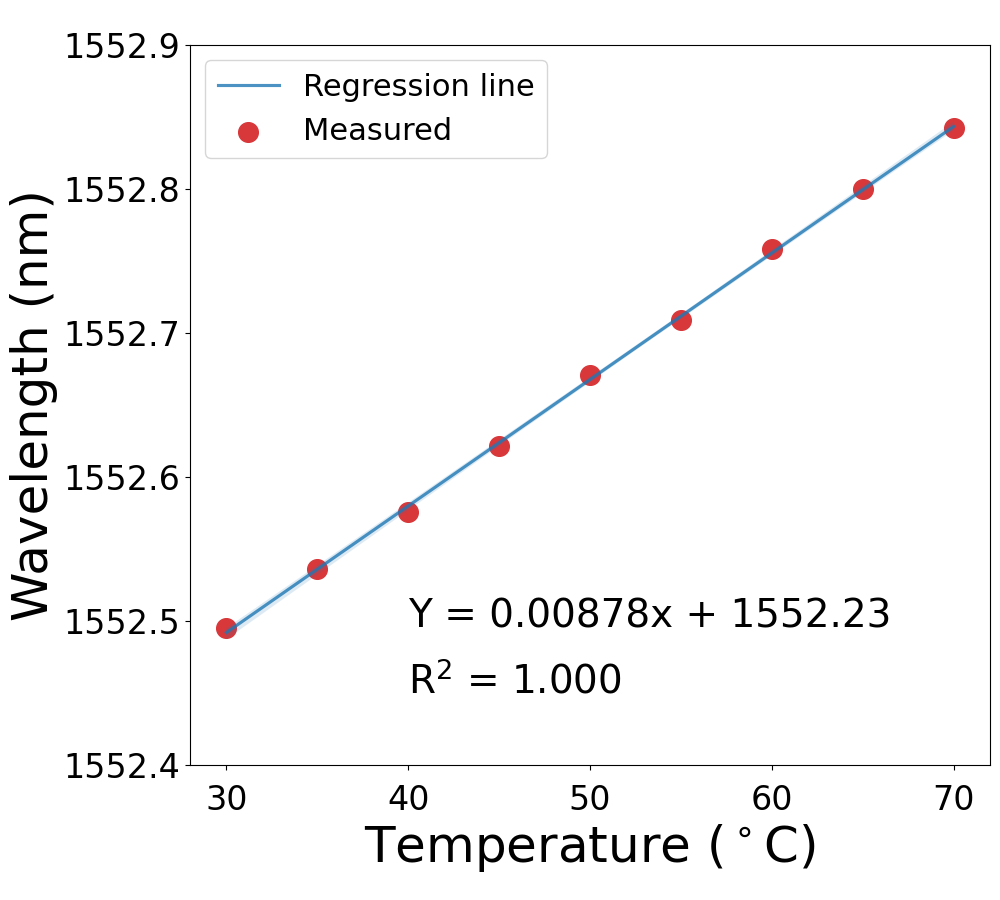}
			\label{fig.core105.fitting_temp}
	\end{minipage}}
	\subfigure[200-$\upmu$m core]{
		\begin{minipage}[t]{0.32\columnwidth}
			\includegraphics[width=1\linewidth]{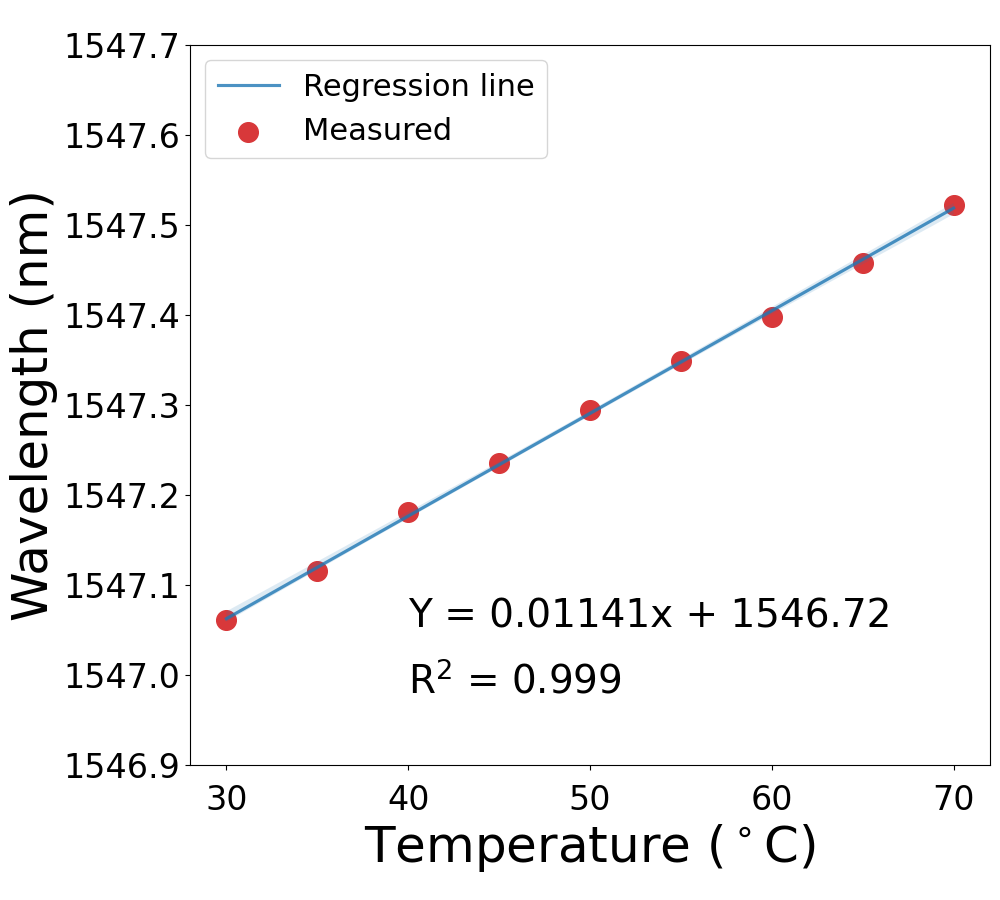}
			\label{fig.core200.fitting_temp}
	\end{minipage}}
	\caption{Temperature measurement results for the MMF with different core diameters. Measured spectral dependence on temperature: (a) 50-$\upmu$m core diameter, (b) 105-$\upmu$m core diameter, and (c) 200-$\upmu$m core diameter. Spectral dip shifts as a function of temperature: (d) 50-$\upmu$m core diameter, (e) 105-$\upmu$m core diameter, and (f) 200-$\upmu$m core diameter.}
	\label{fig.core.temp}
\end{figure}

\begin{figure}[]
	\centering
	\vspace{-0.35cm}
	\subfigtopskip=2pt
	\subfigbottomskip=2pt
	\subfigcapskip=-5pt
	\subfigure[50-$\upmu$m core]{
		\begin{minipage}[t]{0.32\columnwidth}
			\includegraphics[width=1\linewidth]{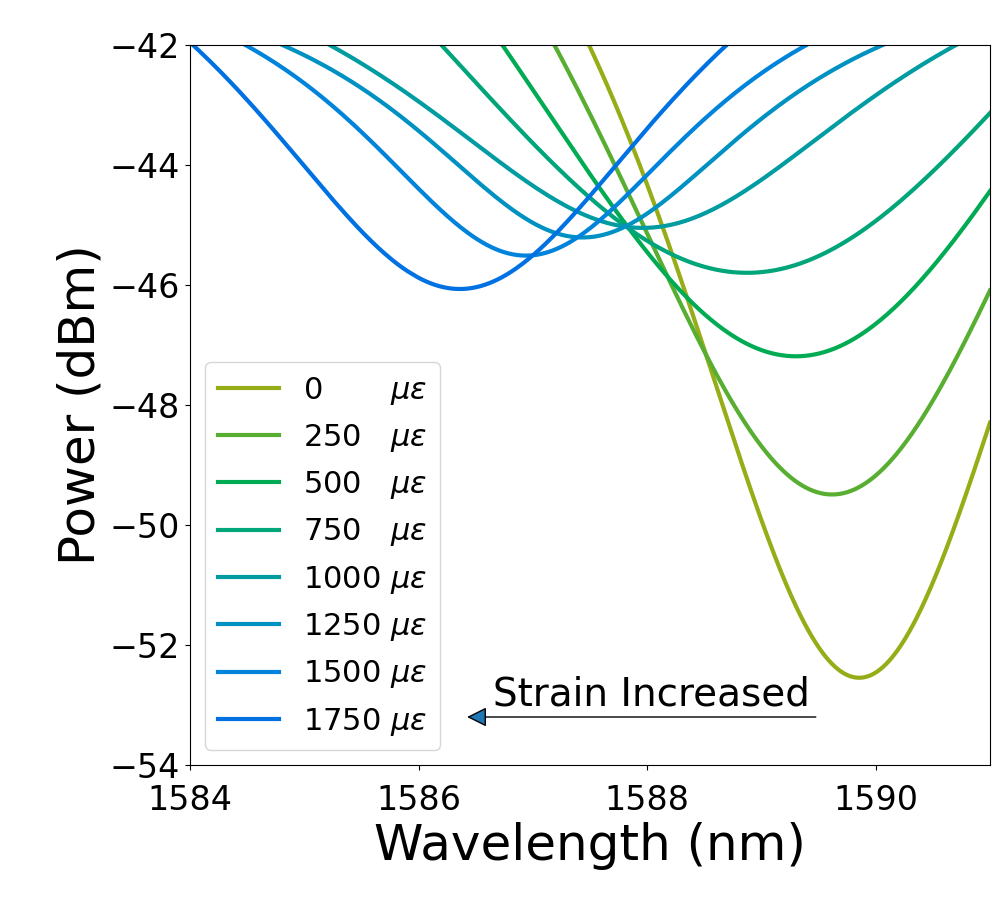}
			\label{fig.core50.wavelength_s}
	\end{minipage}}
	\subfigure[105-$\upmu$m core]{
	\begin{minipage}[t]{0.32\columnwidth}
		\includegraphics[width=1\linewidth]{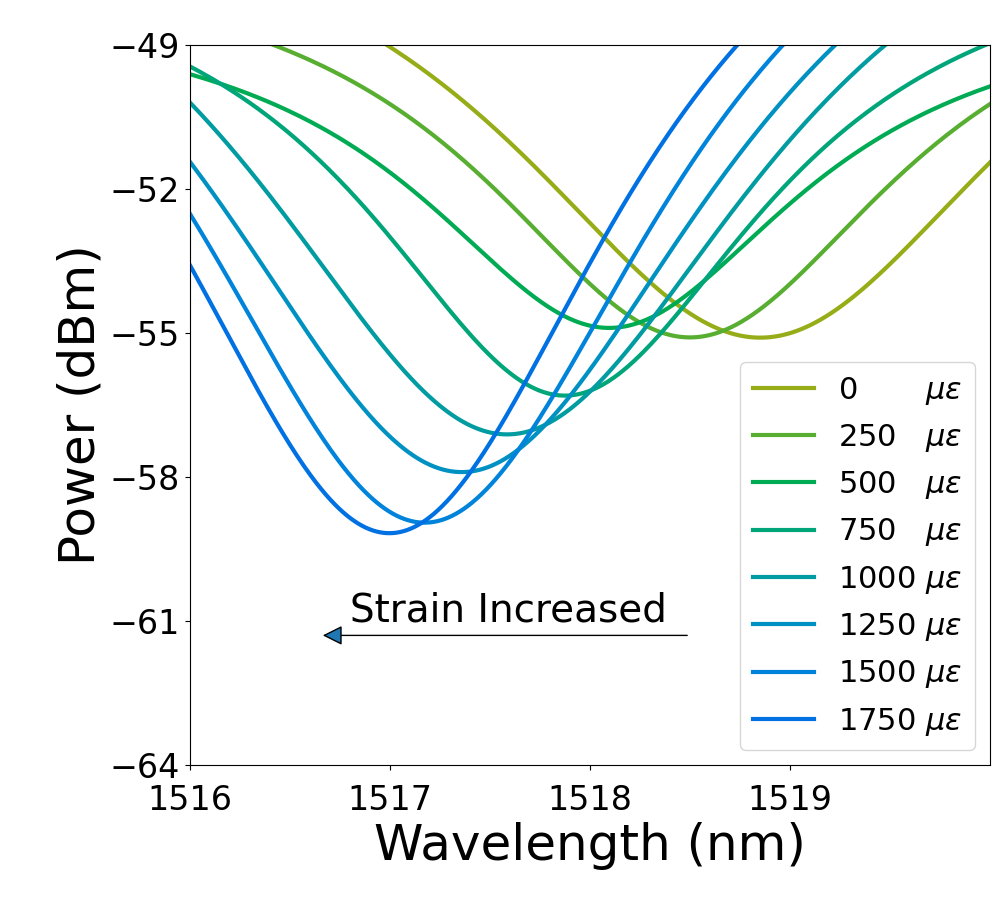}
		\label{fig.core105.wavelength_s}
	\end{minipage}}
	\subfigure[200-$\upmu$m core]{
	\begin{minipage}[t]{0.32\columnwidth}
		\includegraphics[width=1\linewidth]{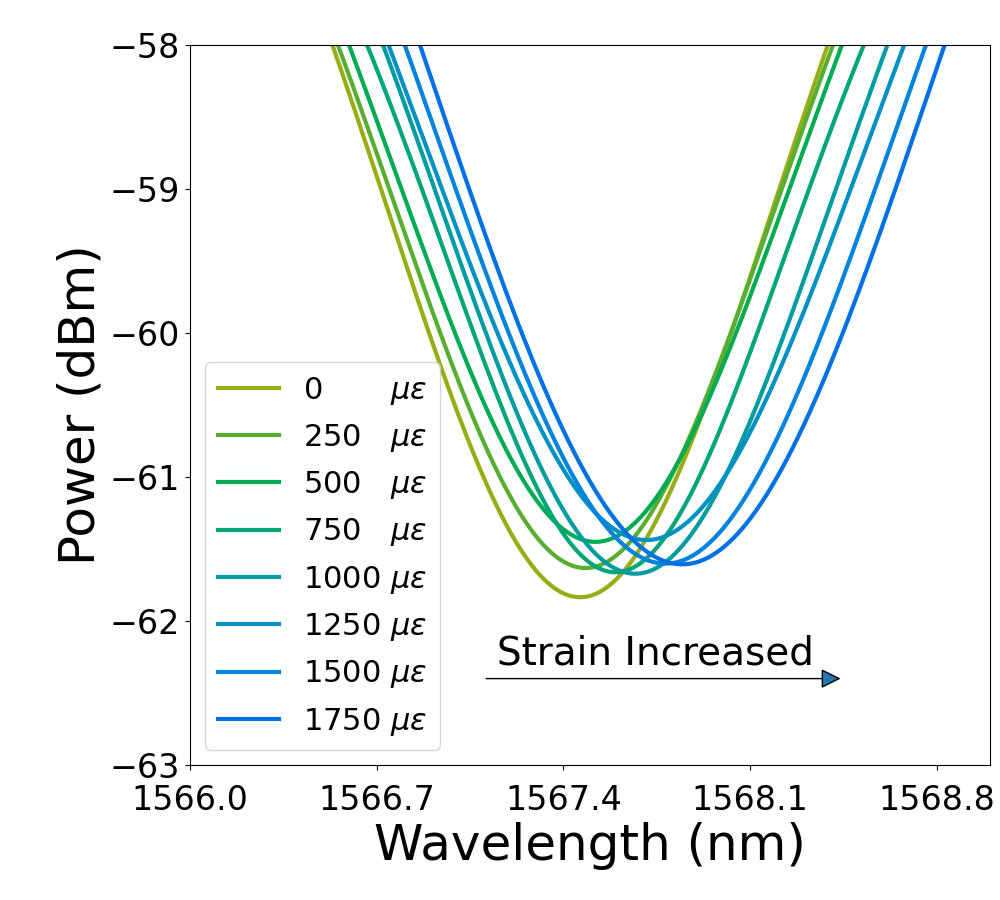}
		\label{fig.core200.wavelength_s}
	\end{minipage}}
\\
	\subfigure[50-$\upmu$m core]{
		\begin{minipage}[t]{0.32\columnwidth}
			\includegraphics[width=1\linewidth]{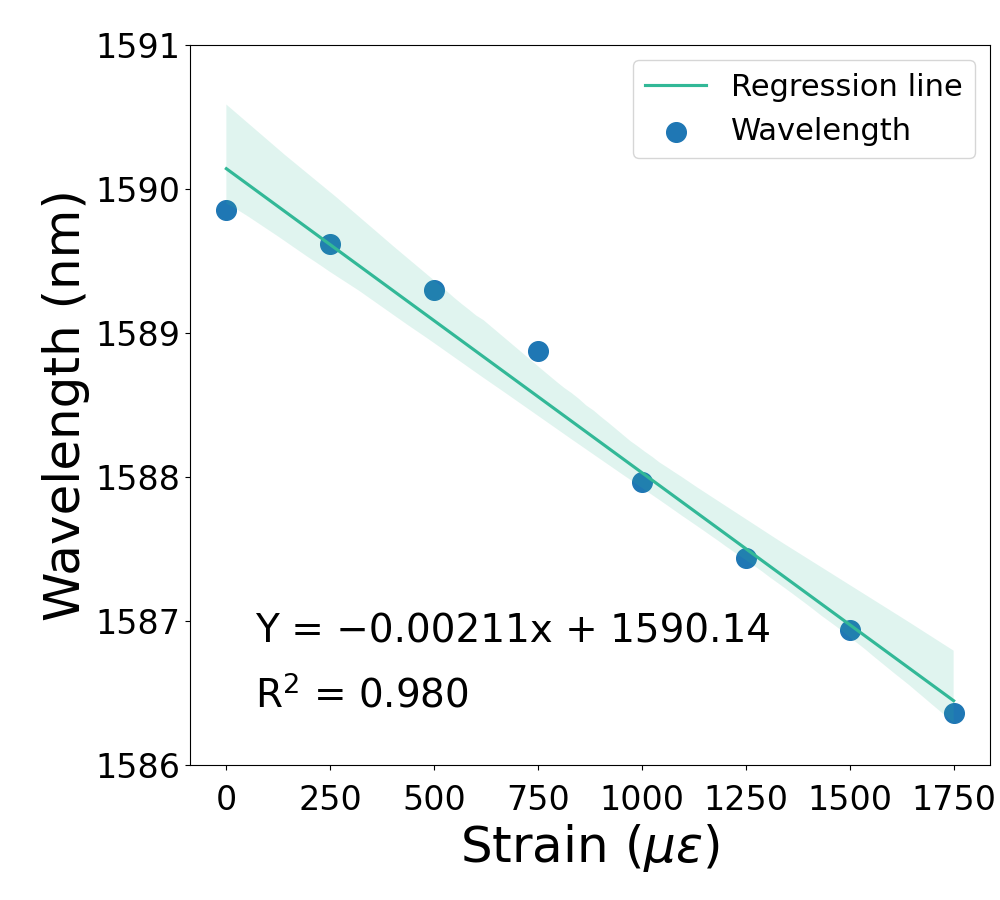}
			\label{fig.core50.fitting_s}
	\end{minipage}}
	\subfigure[105-$\upmu$m core]{
		\begin{minipage}[t]{0.32\columnwidth}
			\includegraphics[width=1\linewidth]{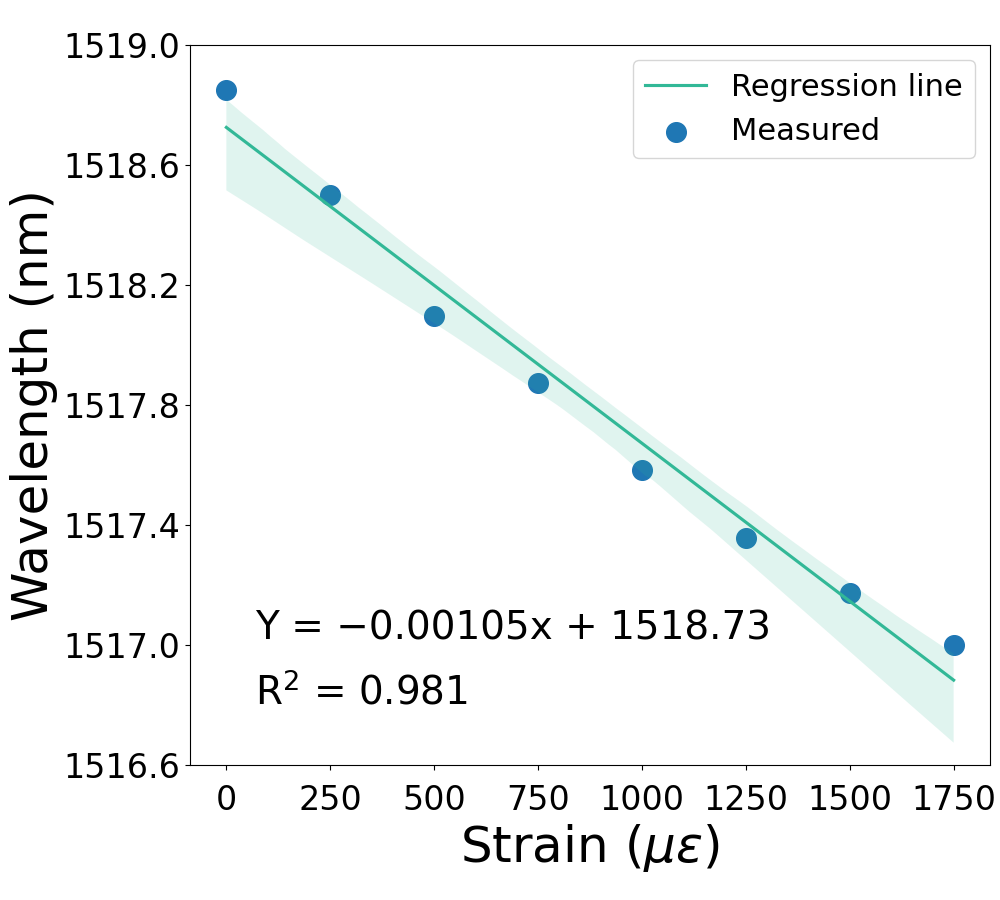}
			\label{fig.core105.fitting_s}
	\end{minipage}}
	\subfigure[200-$\upmu$m core]{
		\begin{minipage}[t]{0.32\columnwidth}
			\includegraphics[width=1\linewidth]{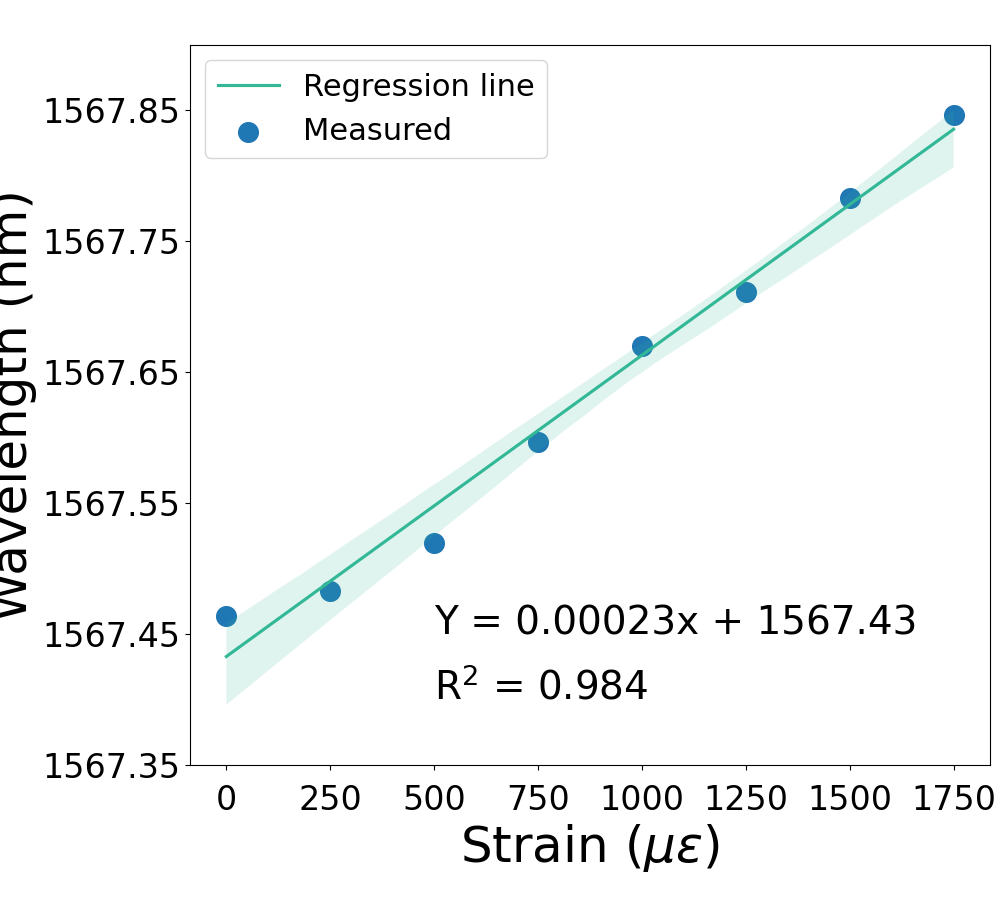}
			\label{fig.core200.fitting_s}
	\end{minipage}}
	\caption{Strain measurement results for the MMF with different core diameters. Measured spectral dependence on strain: (a) 50-$\upmu$m core diameter, (b) 105-$\upmu$m core diameter, and (c) 200-$\upmu$m core diameter. Dip wavelength versus strain: (d) 50-$\upmu$m core diameter, (e) 105-$\upmu$m core diameter, and (f) 200-$\upmu$m core diameter.}
	\label{fig.core.strain}
\end{figure}

The temperature sensitivities of the MMF with different core diameters are illustrated in Fig.~\ref{fig.core.temp}. The measured dependences of the spectral dip on temperature are shown in Fig.~\ref{fig.core50.wavelength_temp}, \ref{fig.core105.wavelength_temp}, and \ref{fig.core200.wavelength_temp} for 50-$\upmu$m, 105-$\upmu$m, and 200-$\upmu$m core diameters, respectively. The spectral dips shift to longer wavelengths when the temperature increases. The resulting wavelength shifts versus temperature are plotted and linearly fitted in Fig.~\ref{fig.core50.fitting_temp}, \ref{fig.core105.fitting_temp}, and \ref{fig.core200.fitting_temp}. The dependences are almost linear, with a coefficient of 8.18~pm/$^{\circ}$C, 8.78~pm/$^{\circ}$C, and 11.41~pm/$^{\circ}$C, respectively. 
In the same way, the strain measurement is presented in Fig.~\ref{fig.core.strain}. With the increasing strain, the spectral dip shifts to shorter wavelengths for 50-$\upmu$m and 105-$\upmu$m but to longer wavelengths for 200-$\upmu$m core diameter, as shown in Fig.~\ref{fig.core50.wavelength_s}, \ref{fig.core105.wavelength_s}, and \ref{fig.core200.wavelength_s}. The corresponding dip wavelength dependence on strain leads to a coefficient of -
2.11~pm/$\upmu\upvarepsilon$, -1.05~pm/$\upmu\upvarepsilon$, and 0.23~pm/$\upmu\upvarepsilon$, as displayed in Fig.~\ref{fig.core50.fitting_s}, \ref{fig.core105.fitting_s}, and \ref{fig.core200.fitting_s}. 

The obtained temperature and strain sensitivities are summarized in Table~\ref{table.core}. The linear regression coefficients ($R^2$) are in the range of 0.980 to 1.000 for both temperature and strain sensing, which is the indication of high linearity. 
This table points out that a larger core diameter of the MMF leads to higher temperature sensitivity but lower strain sensitivity (absolute values). The strain sensitivity dependence on the core diameter also implies that a zero-sensitivity can be realized at a specific core diameter.

\begin{table}[]
	\renewcommand\arraystretch{1.25}
	\centering
	\caption{Temperature and strain sensitivities measured for different core diameters of the MMF.}
	\begin{tabular}{ccccc}
		\toprule
		Fiber Type & Dependence & \makecell[c]{Core\\Diameter} & Sensitivity & $R^2$ \\
		\midrule
		\multirow{6}{*}{\makecell[c]{ \\0.22 NA \\Length = 8 cm}} & \multirow{3}{*}{\makecell[c]{ \\Temperature}} & 50 $\upmu$m & 8.18 pm/$^{\circ}$C & 0.995 \\
		& & 105 $\upmu$m & 8.78 pm/$^{\circ}$C & 1.000\\
		& & 200 $\upmu$m & 11.41 pm/$^{\circ}$C & 0.999\\
		\cmidrule{2-5}
		& \multirow{3}{*}{\makecell[c]{ \\Strain}} & 50 $\upmu$m & -2.11 pm/$\upmu\upvarepsilon$ & 0.980\\
		& & 105 $\upmu$m & -1.05 pm/$\upmu\upvarepsilon$ & 0.981\\
		& & 200 $\upmu$m & 0.23 pm/$\upmu\upvarepsilon$ & 0.984\\
		\bottomrule
	\end{tabular}	
	\label{table.core}
\end{table}

\subsection{Numerical Aperture (NA) Dependence}
The MMFs with a 200-$\upmu$m core diameter have variable numerical apertures: 0.22~NA (FG200LEA, Thorlabs), 0.39~NA (FT200EMT, Thorlabs), and 0.5~NA (FP200ERT, Thorlabs). In this experiment, due to the cladding materials of the MMFs, butt-coupling~\cite{mizuno2010experimental} is used to connect the MMF to the SMFs. Due to the configuration of the fiber sensor, a 20-cm-long MMF section is used for the measurement. 

\begin{figure}[]
	\centering
	\vspace{-0.35cm}
	\subfigtopskip=2pt
	\subfigbottomskip=2pt
	\subfigcapskip=-5pt
	\subfigure[0.22 NA]{
		\begin{minipage}[t]{0.32\columnwidth}
			\includegraphics[width=1\linewidth]{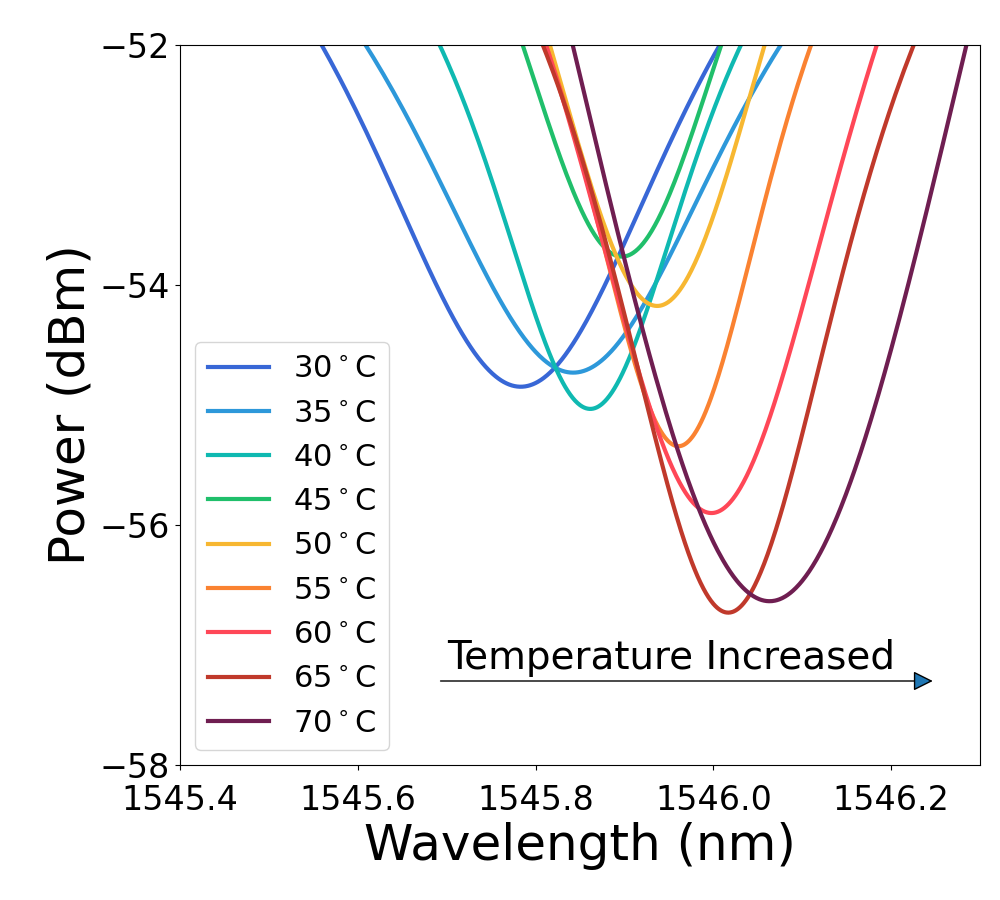}
			\label{fig.na22.wavelength_temp}
	\end{minipage}}
	\subfigure[0.39 NA]{
	\begin{minipage}[t]{0.32\columnwidth}
		\includegraphics[width=1\linewidth]{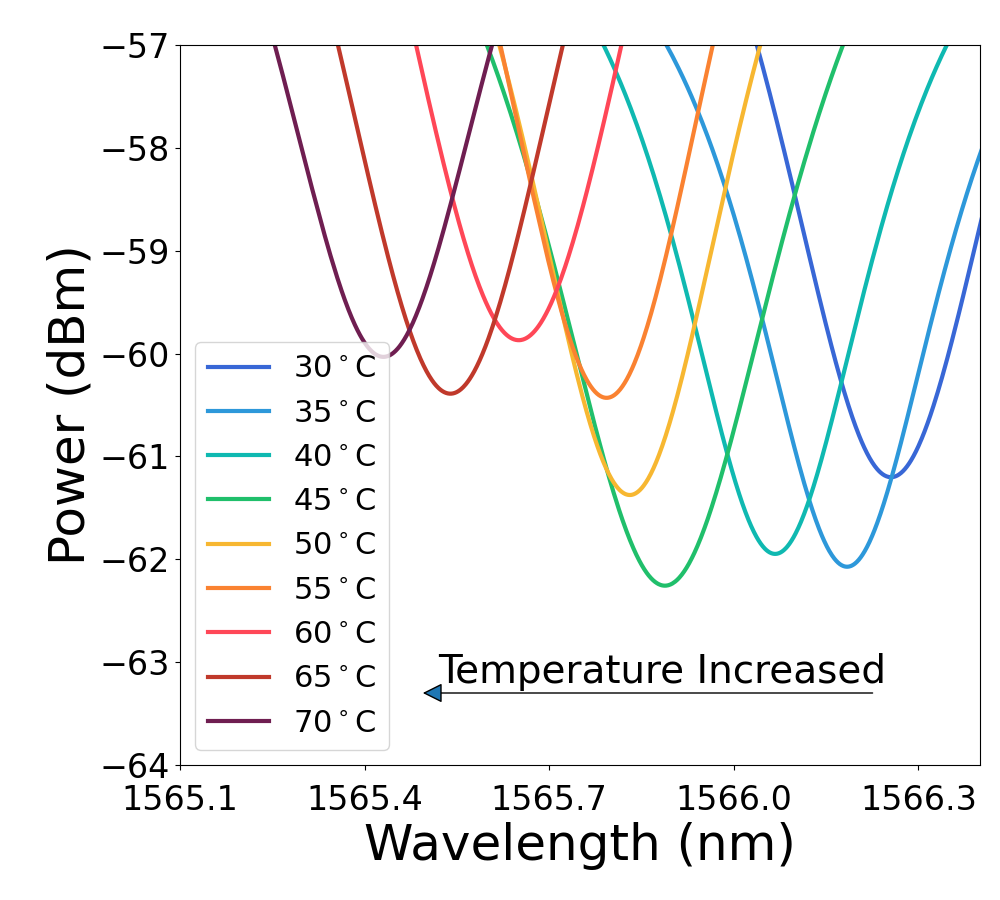}
		\label{fig.na39.wavelength_temp}
	\end{minipage}}
	\subfigure[0.5 NA]{
	\begin{minipage}[t]{0.32\columnwidth}
		\includegraphics[width=1\linewidth]{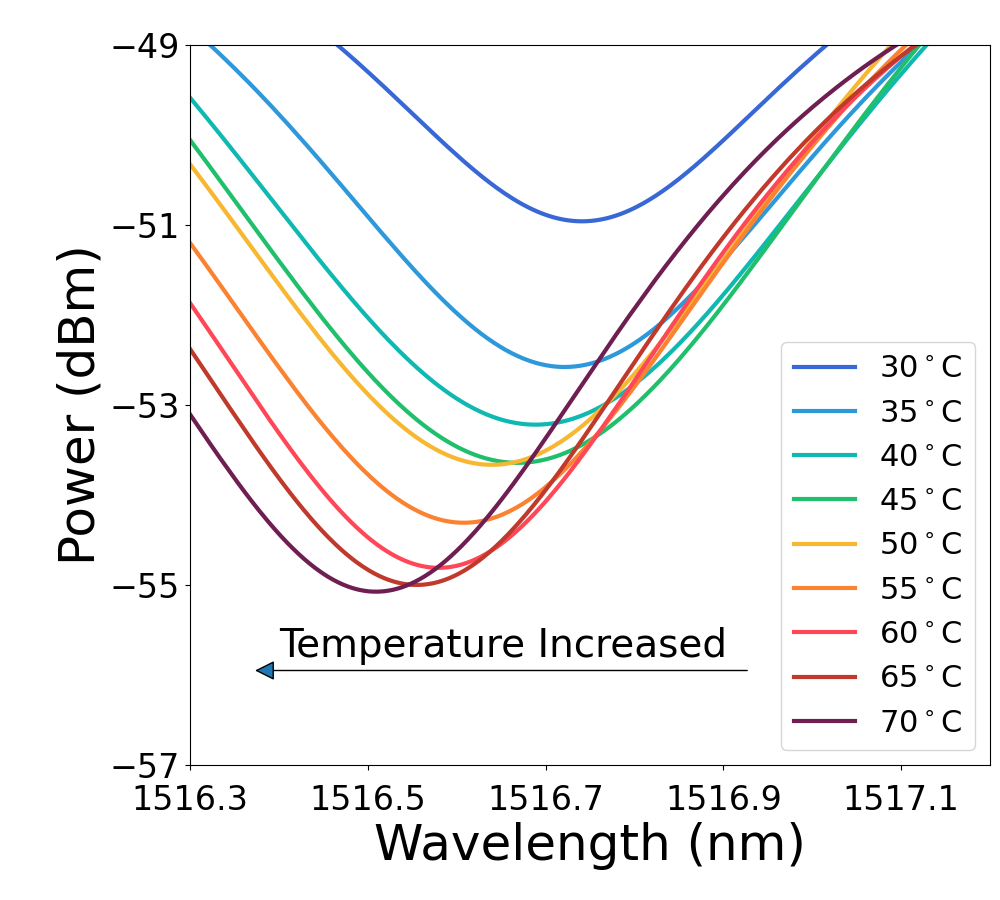}
		\label{fig.na5.wavelength_temp}
	\end{minipage}}
\\
	\subfigure[0.22 NA]{
		\begin{minipage}[t]{0.32\columnwidth}
			\includegraphics[width=1\linewidth]{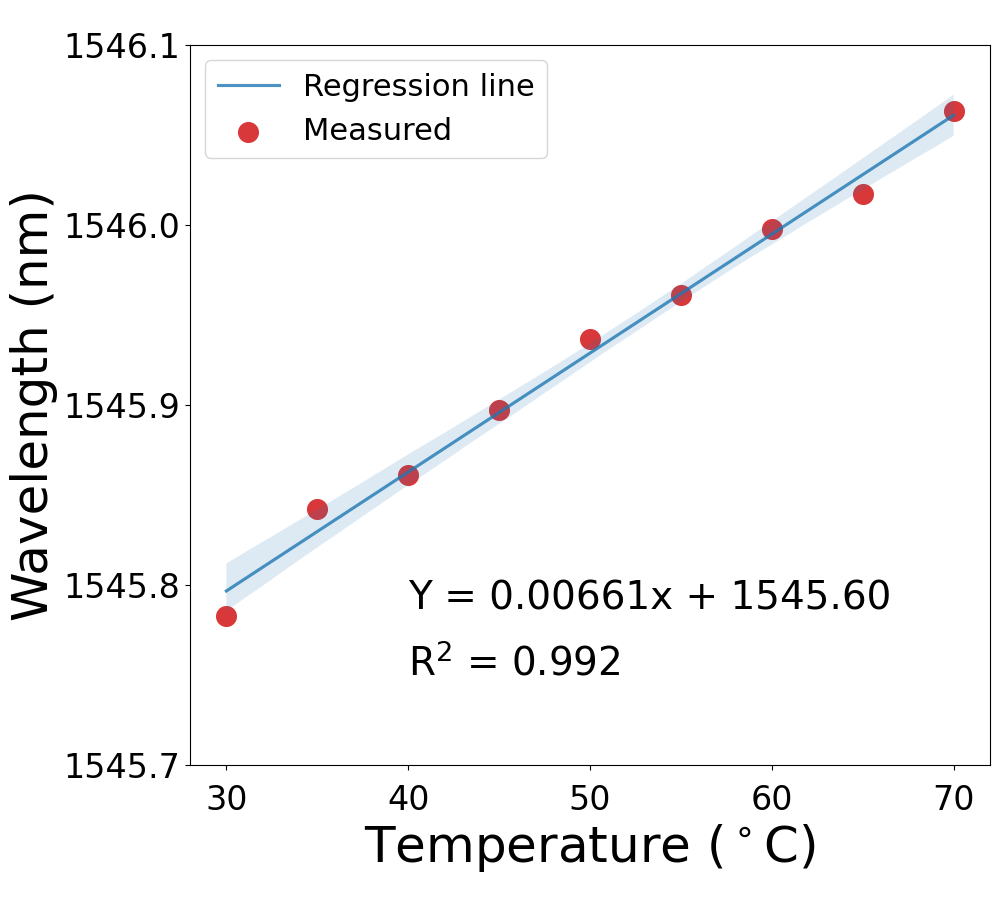}
			\label{fig.na22.fitting_temp}
	\end{minipage}}
	\subfigure[0.39 NA]{
		\begin{minipage}[t]{0.32\columnwidth}
			\includegraphics[width=1\linewidth]{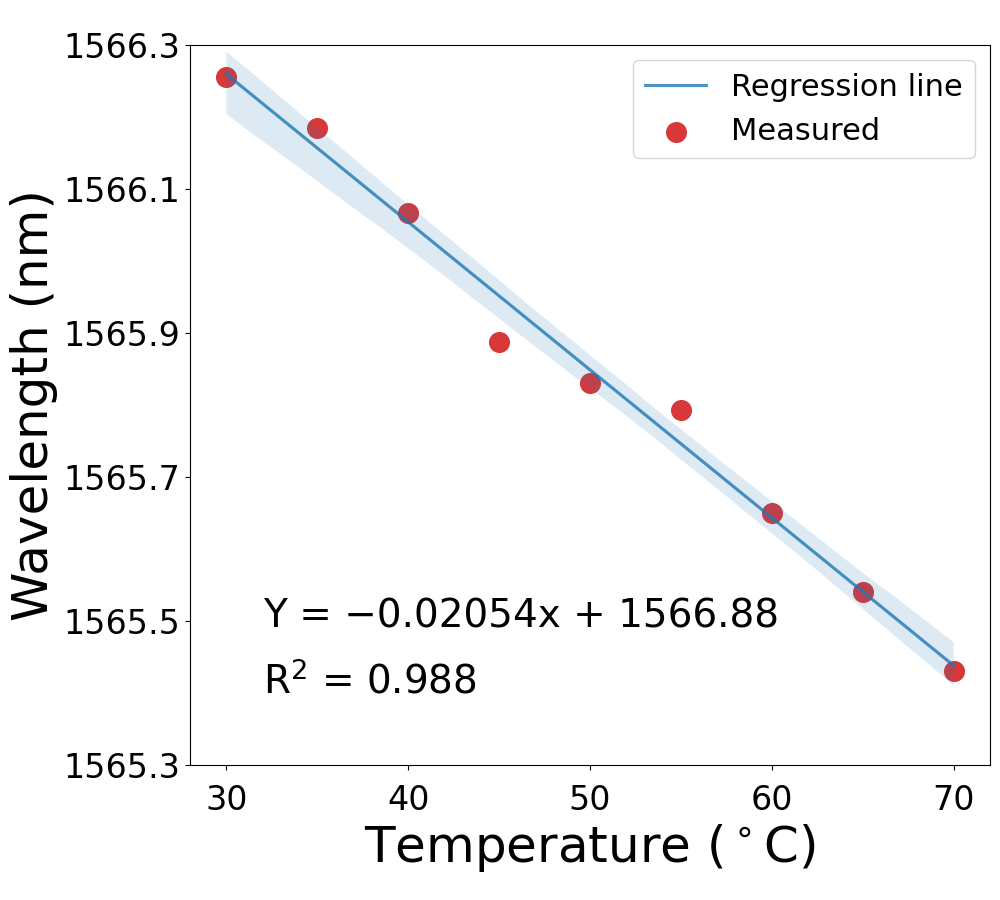}
			\label{fig.na39.fitting_temp}
	\end{minipage}}
	\subfigure[0.5 NA]{
		\begin{minipage}[t]{0.32\columnwidth}
			\includegraphics[width=1\linewidth]{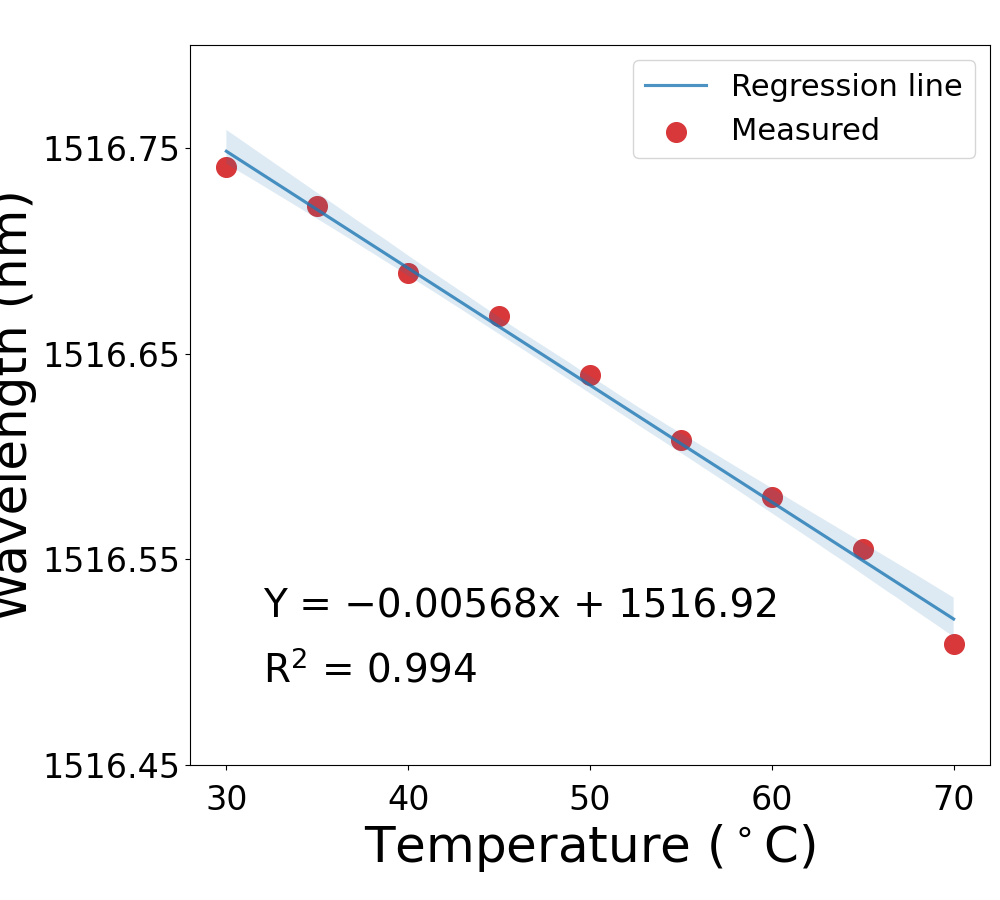}
			\label{fig.na5.fitting_temp}
	\end{minipage}}
	\caption{Temperature measurement results for the MMF with different NAs. Measured spectral dependence on temperature: (a) 0.22 NA, (b) 0.39 NA, and (c) 0.5 NA. Spectral dip shifts as a function of temperature: (d) 0.22 NA, (e) 0.39 NA, and (f) 0.5 NA.}
	\label{fig.NA.temp}
\end{figure}

\begin{figure}[]
	\centering
	\subfigtopskip=2pt
	\subfigbottomskip=2pt
	\subfigcapskip=-5pt
	\subfigure[0.22 NA]{
		\begin{minipage}[t]{0.32\columnwidth}
			\includegraphics[width=1\linewidth]{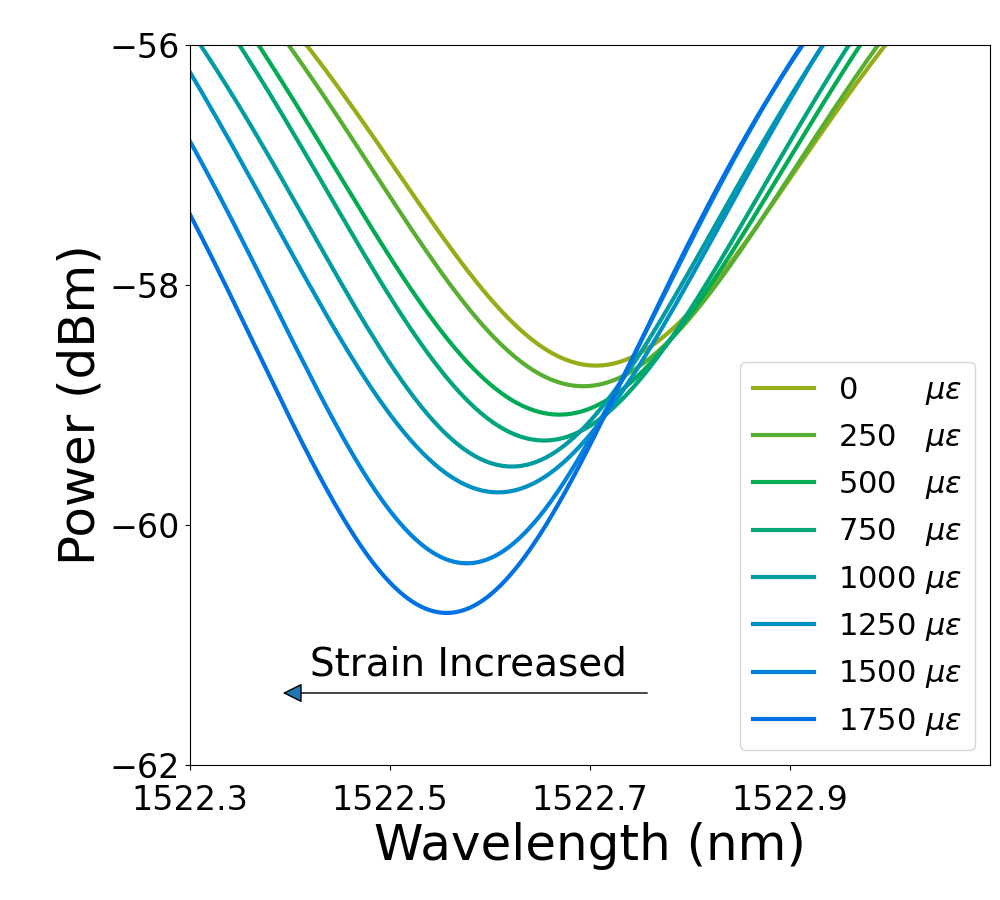}
			\label{fig.na22.wavelength_strain}
	\end{minipage}}
	\subfigure[0.39 NA]{
	\begin{minipage}[t]{0.32\columnwidth}
		\includegraphics[width=1\linewidth]{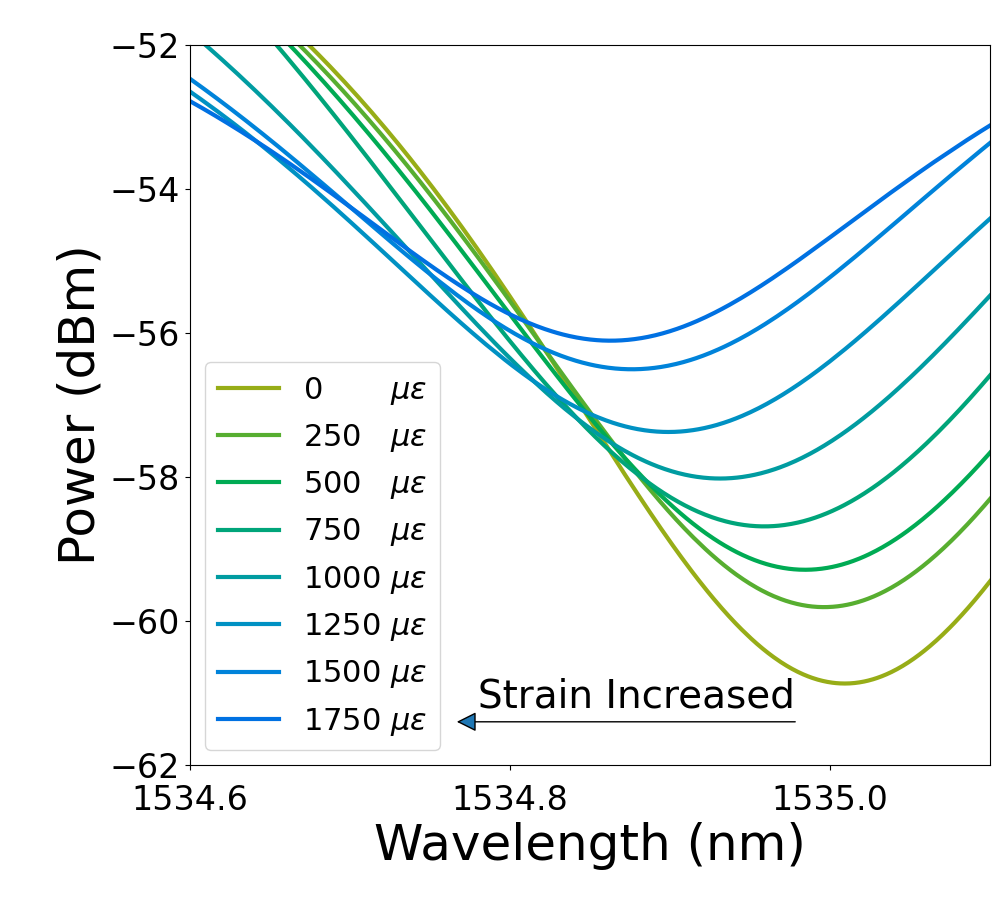}
		\label{fig.na39.wavelength_strain}
	\end{minipage}}
	\subfigure[0.5 NA]{
	\begin{minipage}[t]{0.32\columnwidth}
		\includegraphics[width=1\linewidth]{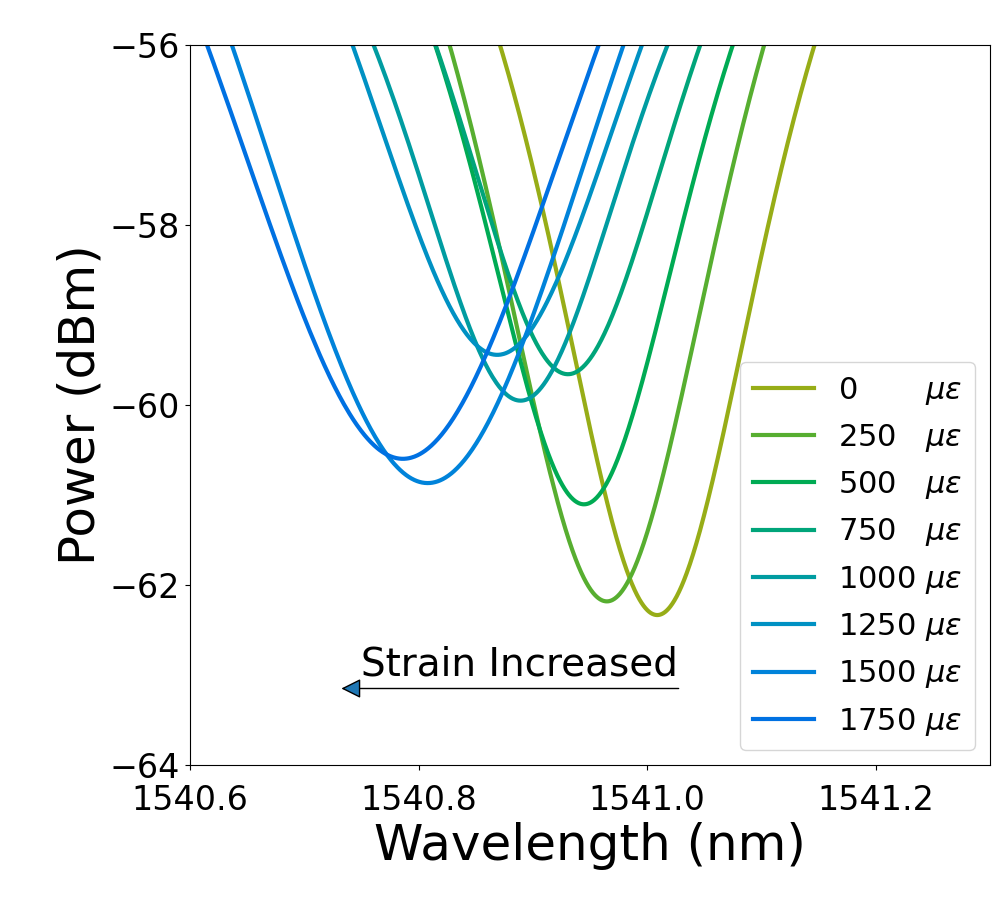}
		\label{fig.na5.wavelength_strain}
	\end{minipage}}
\\
	\subfigure[0.22 NA]{
		\begin{minipage}[t]{0.32\columnwidth}
			\includegraphics[width=1\linewidth]{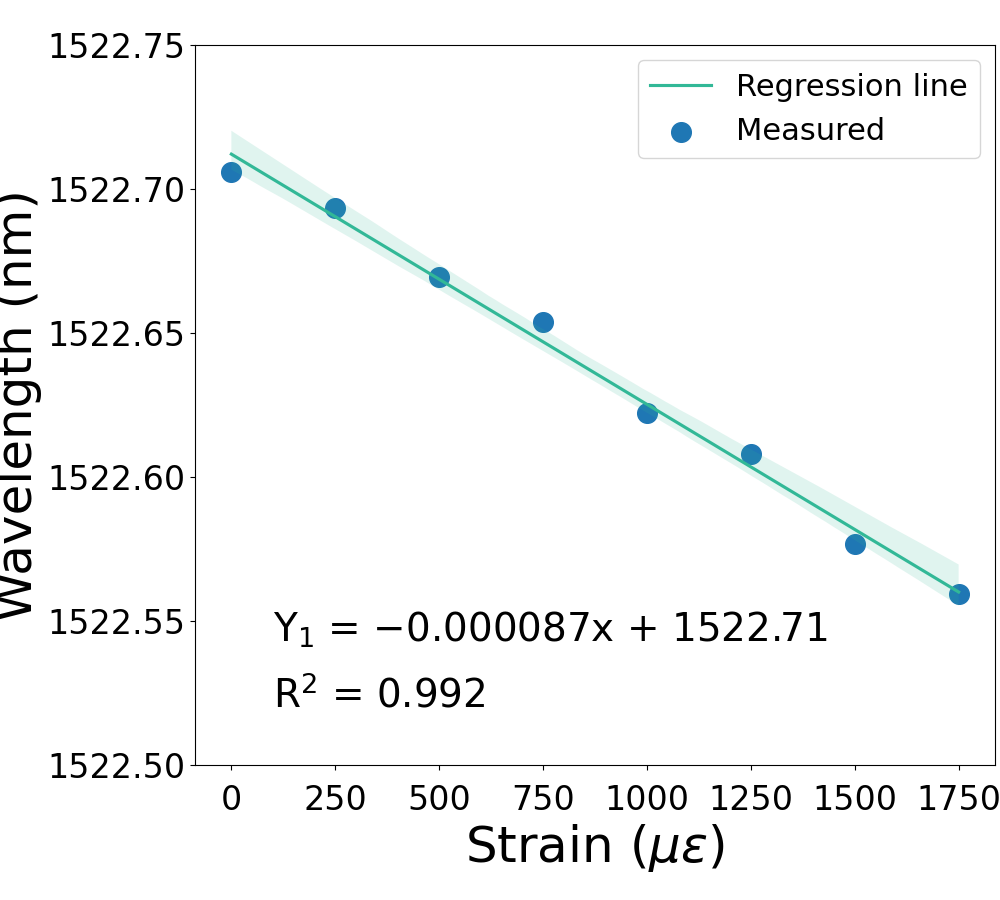}
			\label{fig.na22.fitting_strain}
	\end{minipage}}
	\subfigure[0.39 NA]{
		\begin{minipage}[t]{0.32\columnwidth}
			\includegraphics[width=1\linewidth]{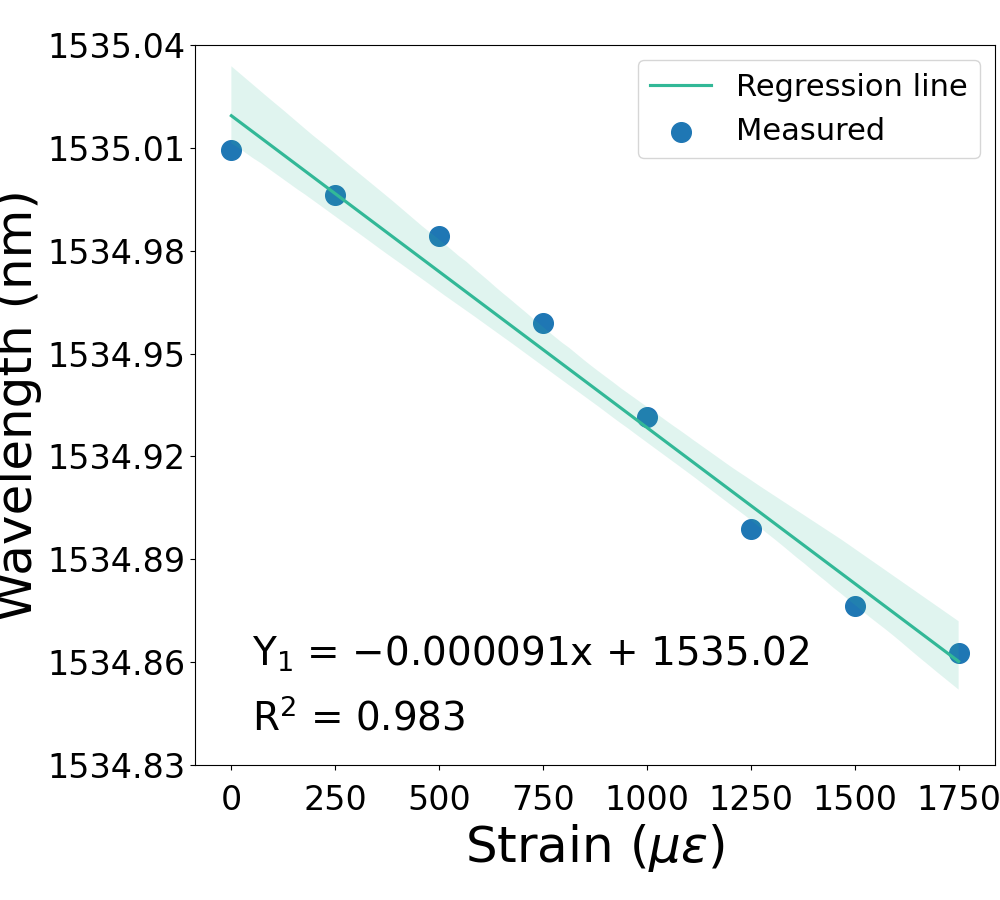}
			\label{fig.na39.fitting_strain}
	\end{minipage}}
	\subfigure[0.5 NA]{
		\begin{minipage}[t]{0.32\columnwidth}
			\includegraphics[width=1\linewidth]{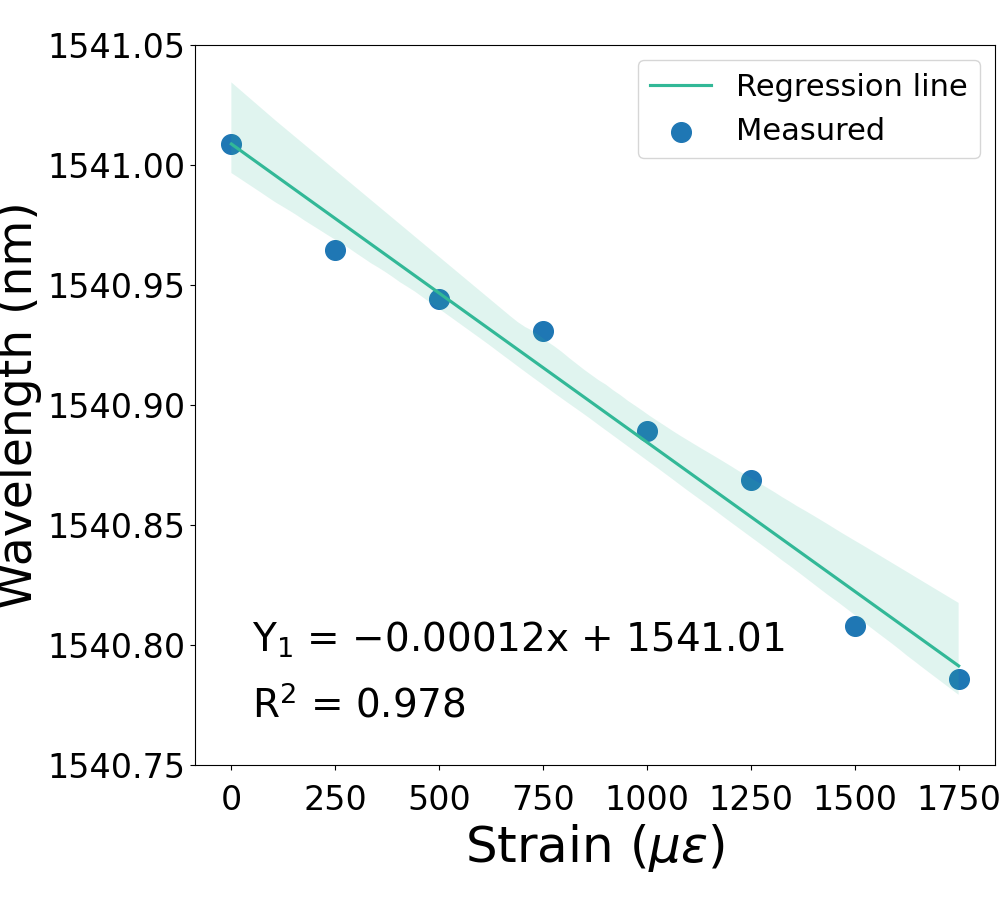}
			\label{fig.na5.fitting_strain}
	\end{minipage}}
	\caption{Strain measurement results for the MMF with different NAs. (a) 0.22 NA, (b) 0.39 NA, and (c) 0.5 NA. Dip wavelength versus strain: (d) 0.22 NA, (e) 0.39 NA, and (f) 0.5 NA.}
	\label{fig.NA.strain}
\end{figure}

The temperature sensitivity of the MMF with different NAs is demonstrated in Fig.~\ref{fig.NA.temp}. Figure~\ref{fig.na22.wavelength_temp} shows a clear redshift of the wavelength dips with the temperature increases for the MMF with 0.22~NA, while Fig.~\ref{fig.na39.wavelength_temp} and Fig.~\ref{fig.na5.wavelength_temp} show a blueshift for the MMF with 0.39 and 0.5~NA, respectively. Figure~\ref{fig.na22.fitting_temp}, \ref{fig.na39.fitting_temp}, and \ref{fig.na5.fitting_temp} present the corresponding measured temperature sensitivities of 6.61~pm/$^{\circ}$C, -20.54~pm/$^{\circ}$C, and -5.68~pm/$^{\circ}$C. 
The strain sensitivity of the MMF with different NAs is illustrated in Fig.~\ref{fig.NA.strain}. The clear blueshifts of the wavelength dips are distinguished with strain applying for the MMF with 0.22, 0.39, and 0.5~NA, as shown in Fig.~\ref{fig.na22.wavelength_strain}, ~\ref{fig.na39.wavelength_strain}, and ~\ref{fig.na5.wavelength_strain}. The related measured strain sensitivities are -0.087~pm/$\upmu\upvarepsilon$, -0.091~pm/$\upmu\upvarepsilon$, and -0.120~pm/$\upmu\upvarepsilon$, as plotted in Fig.~\ref{fig.na22.fitting_strain}, ~\ref{fig.na39.fitting_strain}, and ~\ref{fig.na5.fitting_strain}. 

The obtained temperature and strain dependences on the NA of the MMF are summarized in Table~\ref{table.NA}, in which the $R^2$ values also illustrate high linearity. It expresses that the temperature sensitivity is independent of the NA of the MMF, while the absolute value of strain sensitivity increases with larger NA. 

\begin{table}[]
	\renewcommand\arraystretch{1.25}
	\centering
	\caption{Temperature and strain sensitivities measured for different numerical apertures (NAs) of the MMF.}
	\begin{tabular}{ccccc}
		\toprule
		Fiber Type & Dependence & NA & Sensitivity & $R^2$ \\
		\midrule
		\multirow{6}{*}{\makecell[l]{ \\Core Diameter\\ = 200 $\upmu$m \\Length = 20 cm}} & \multirow{3}{*}{\makecell[c]{ \\Temperature}} & 0.22 & 6.61 pm/$^{\circ}$C & 0.992 \\
		& & 0.39 & -20.54 pm/$^{\circ}$C & 0.988\\
		& & 0.5 & -5.68 pm/$^{\circ}$C & 0.994\\
		\cmidrule{2-5}
		& \multirow{3}{*}{\makecell[c]{ \\Strain}} & 0.22 & -0.087 pm/$\upmu\upvarepsilon$ & 0.992\\
		& & 0.39 & -0.091 pm/$\upmu\upvarepsilon$ & 0.983\\
		& & 0.5 & -0.120 pm/$\upmu\upvarepsilon$ & 0.978\\
		\bottomrule
	\end{tabular}	
	\label{table.NA}
\end{table}

\subsection{Length Dependence}
As described before, the principle of the SMS fiber sensor has an assumption where only two modes of the MMF (the fundamental and the first-order higher modes) are excited. However, other modes also need to be considered in the experiments, and it leads to a discussion of whether the temperature and strain sensitivities are dependent on the MMF lengths~\cite{YosukeMIZUNO2018}. 
As the scope of this work is to investigate the characteristics of standard MMFs, the 50-$\upmu$m core diameter MMF with a 0.22~NA is used. 

\begin{figure}[]
	\vspace{-2cm}
	\subfigtopskip=2pt
	\subfigbottomskip=2pt
	\subfigcapskip=-5pt
	\subfigure[2 cm]{
		\begin{minipage}[t]{0.32\columnwidth}
			\includegraphics[width=1\linewidth]{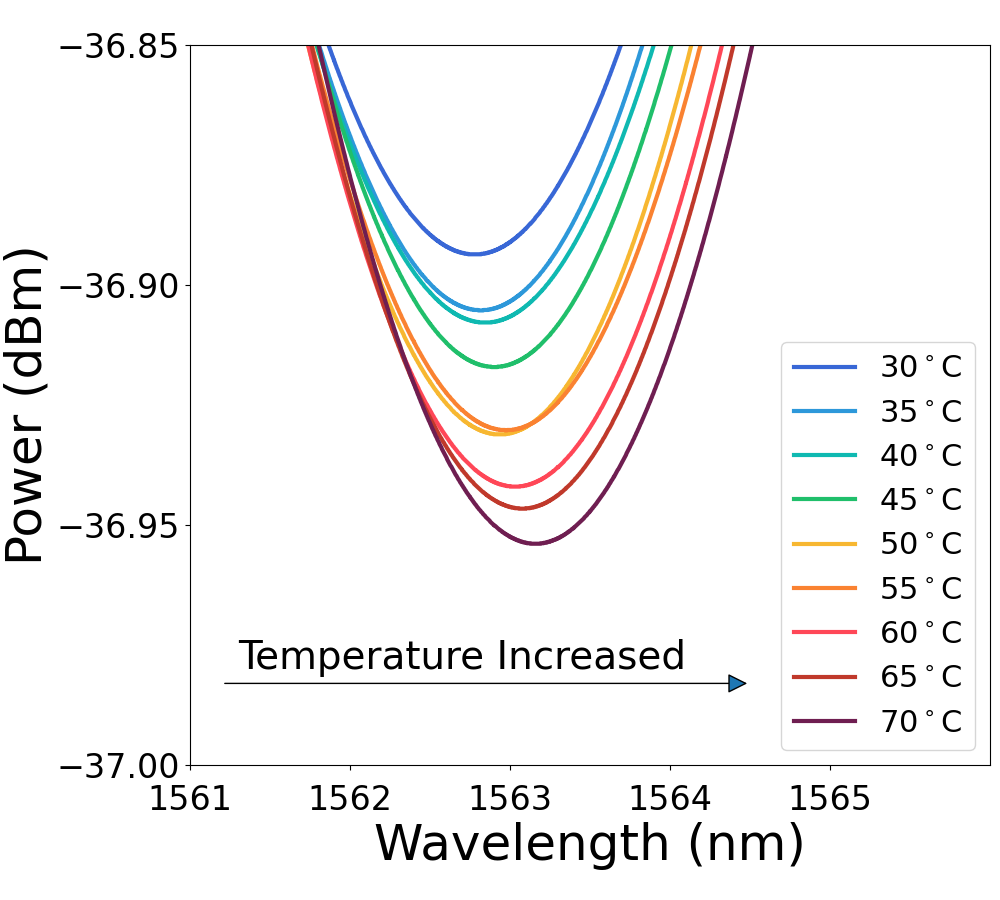}
			\label{fig.2cm.wavelength_t}
	\end{minipage}}
	\subfigure[4 cm]{
	\begin{minipage}[t]{0.32\columnwidth}
		\includegraphics[width=1\linewidth]{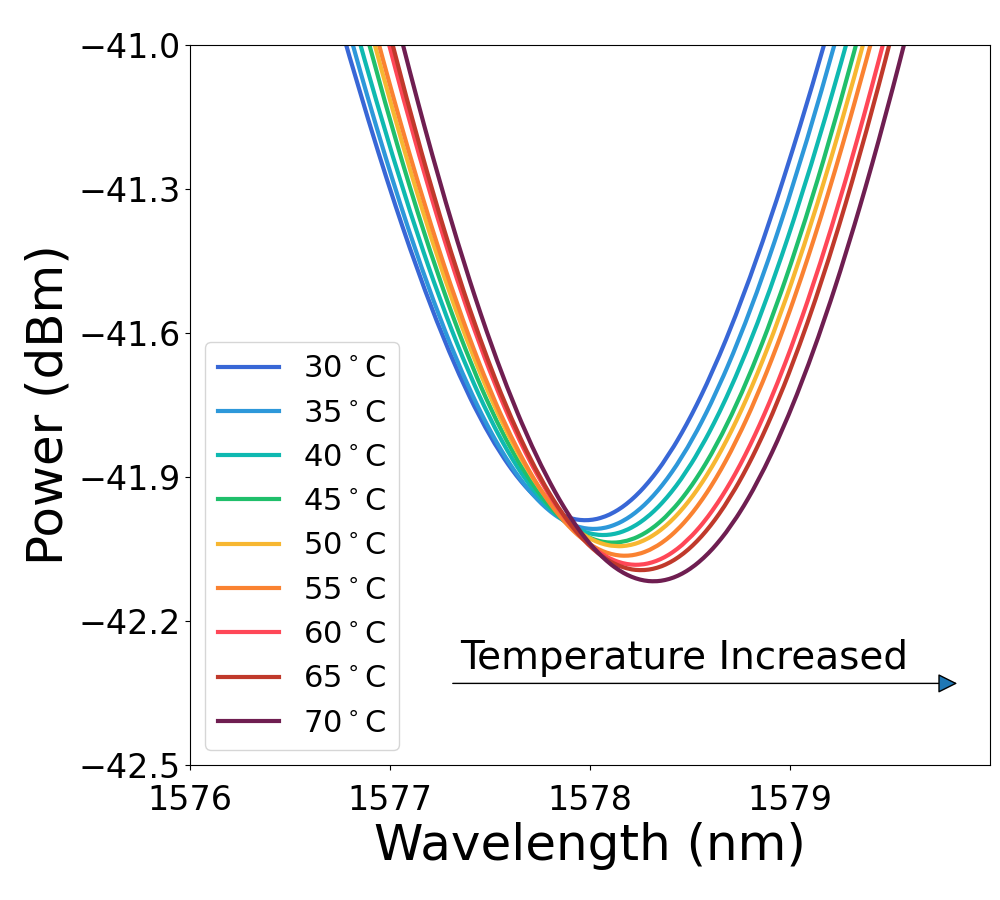}
		\label{fig.4cm.wavelength_t}
	\end{minipage}}
	\subfigure[8 cm]{
	\begin{minipage}[t]{0.32\columnwidth}
		\includegraphics[width=1\linewidth]{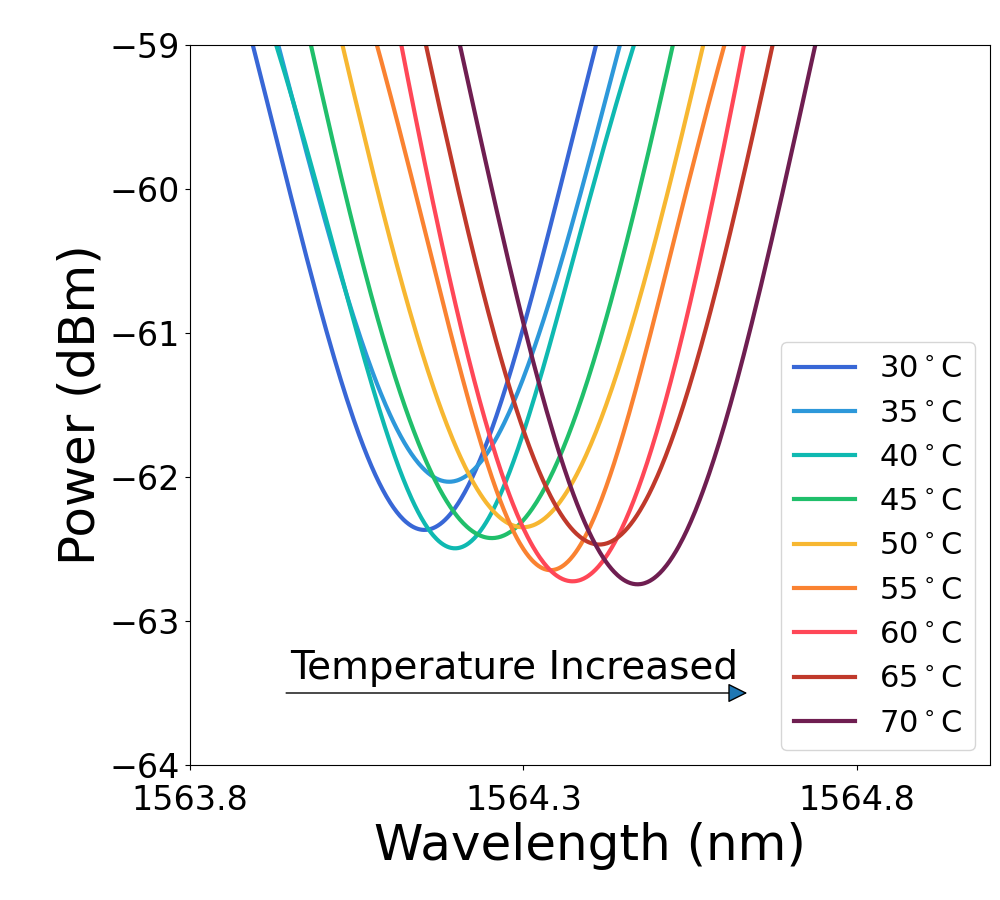}
		\label{fig.8cm.wavelength_t}
	\end{minipage}}
\\
	\subfigure[12 cm]{
	\begin{minipage}[t]{0.32\columnwidth}
		\flushleft
		\includegraphics[width=1\linewidth]{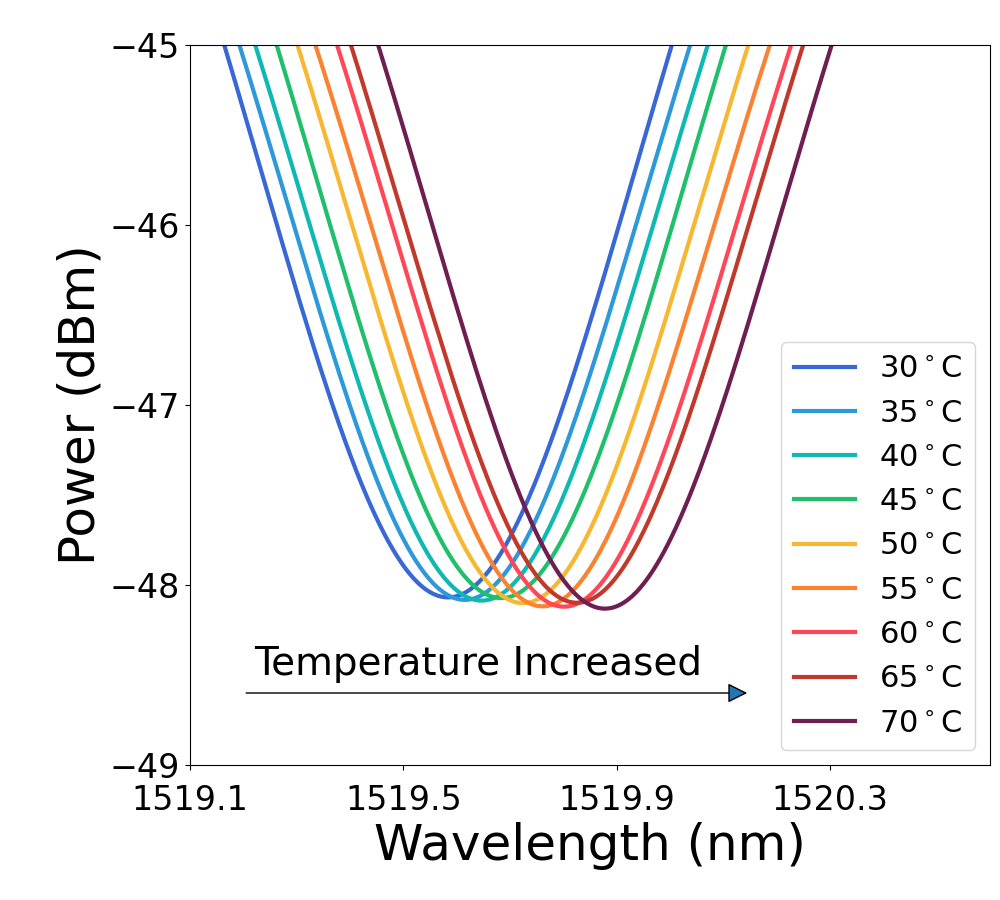}
		\label{fig.12cm.wavelength_t}
	\end{minipage}}
	\subfigure[16 cm]{
	\begin{minipage}[t]{0.32\columnwidth}
		\flushleft
		\includegraphics[width=1\linewidth]{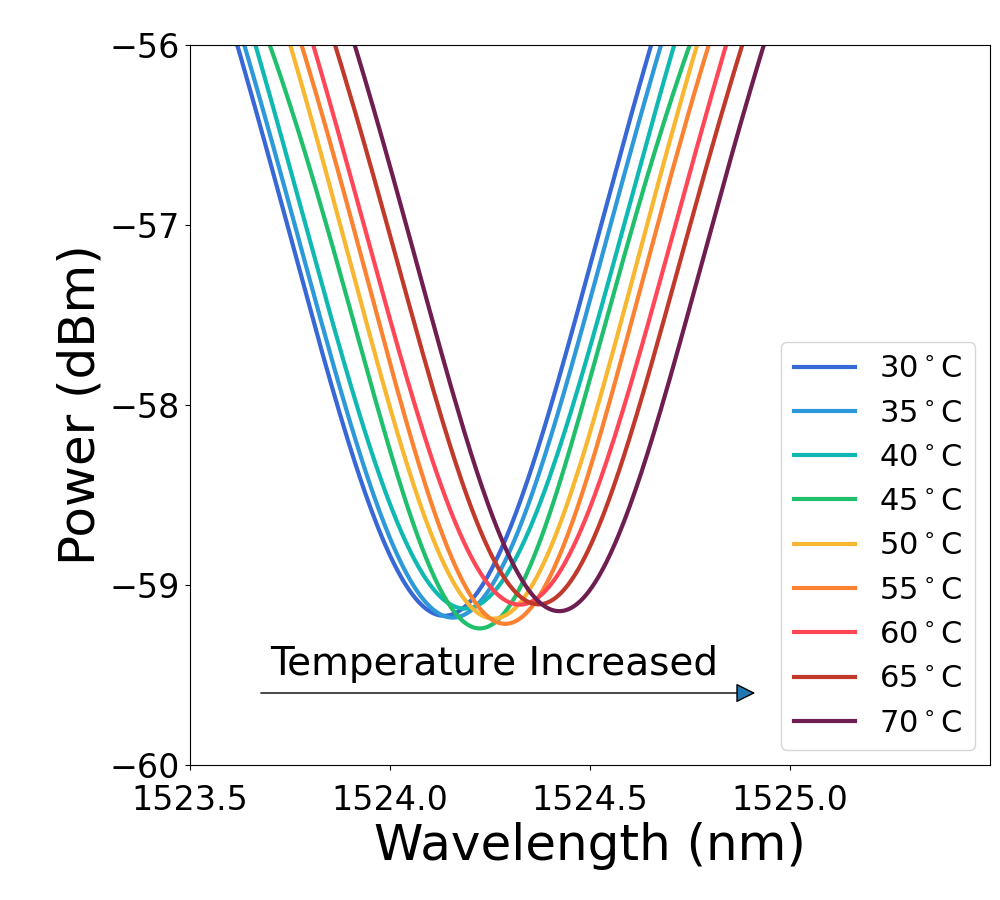}
		\label{fig.16cm.wavelength_t}
	\end{minipage}}
\\
	\subfigure[2 cm]{
		\begin{minipage}[t]{0.32\columnwidth}
			\includegraphics[width=1\linewidth]{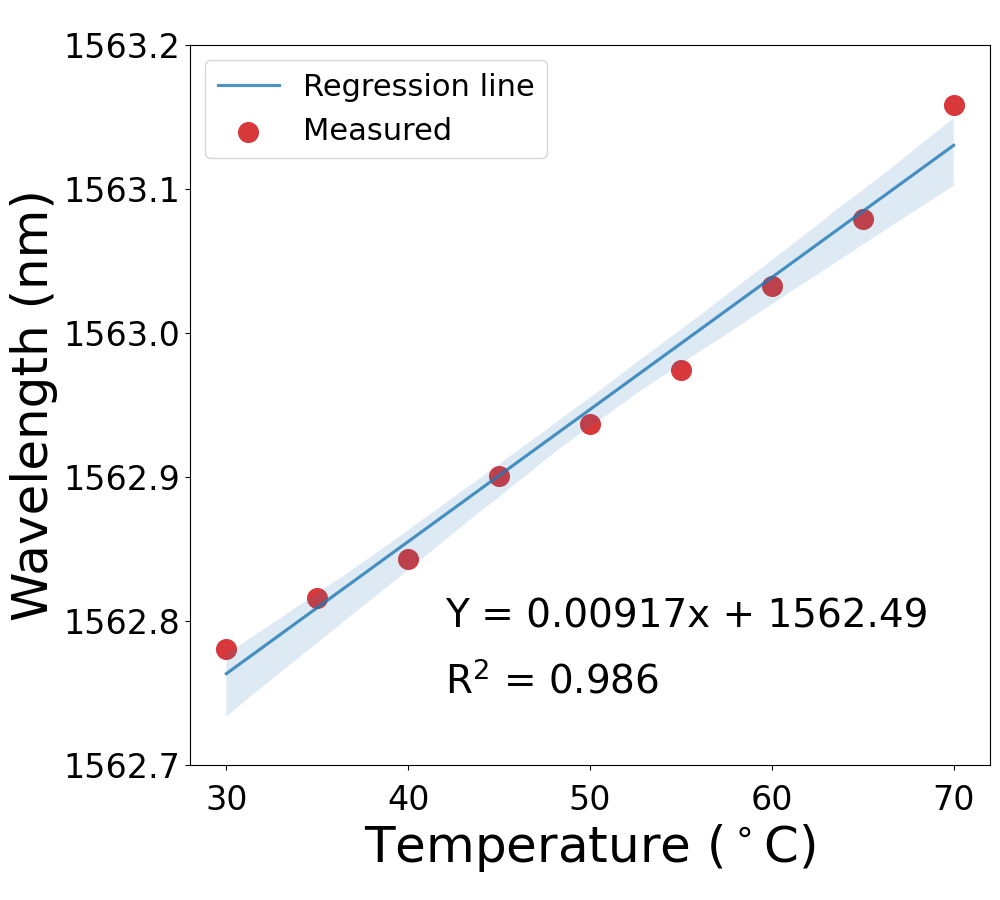}
			\label{fig.2cm.fitting_t}
	\end{minipage}}
	\subfigure[4 cm]{
		\begin{minipage}[t]{0.32\columnwidth}
			\includegraphics[width=1\linewidth]{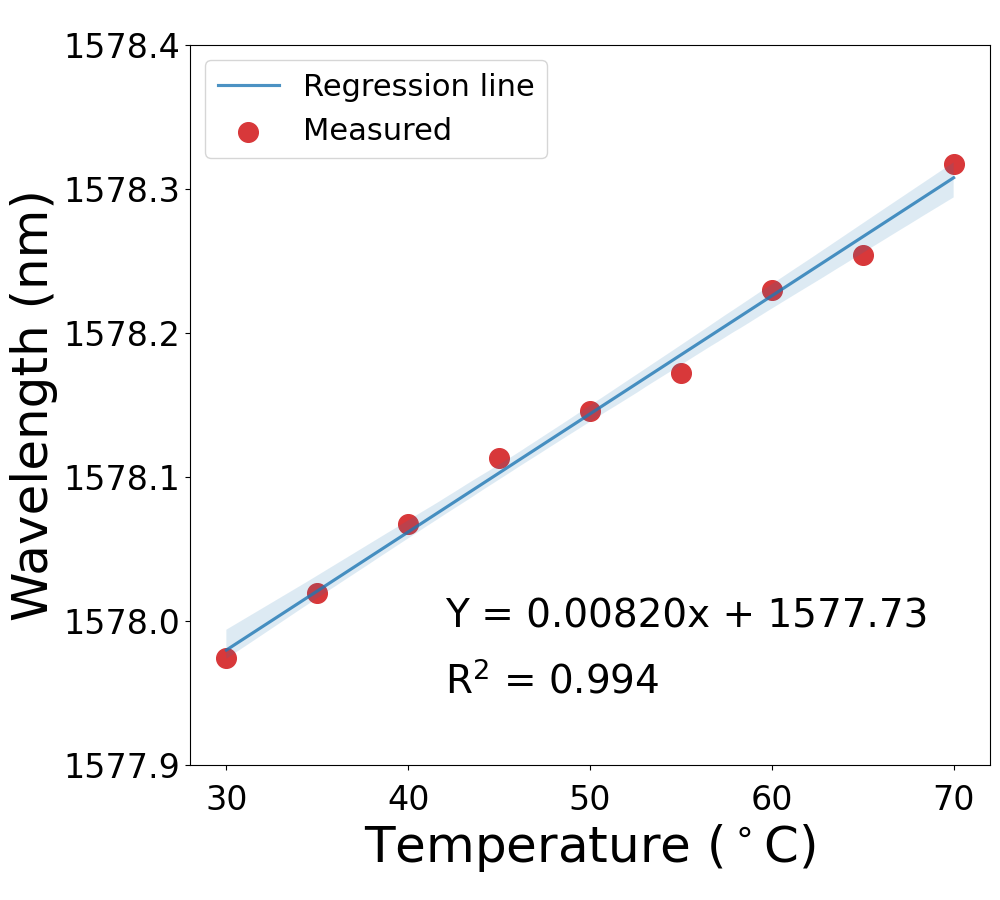}
			\label{fig.4cm.fitting_t}
	\end{minipage}}
	\subfigure[8 cm]{
		\begin{minipage}[t]{0.32\columnwidth}
			\includegraphics[width=1\linewidth]{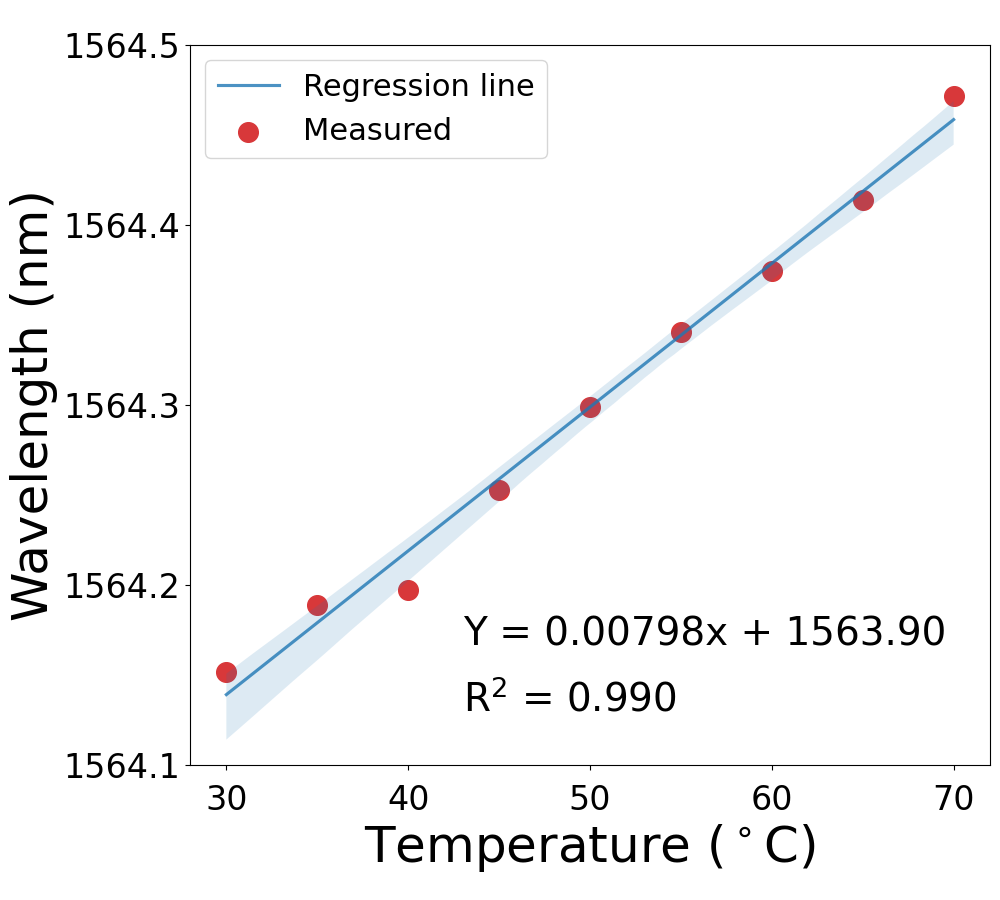}
			\label{fig.8cm.fitting_t}
	\end{minipage}}
\\
	\subfigure[12 cm]{
		\begin{minipage}[t]{0.32\columnwidth}
			\flushleft
			\includegraphics[width=1\linewidth]{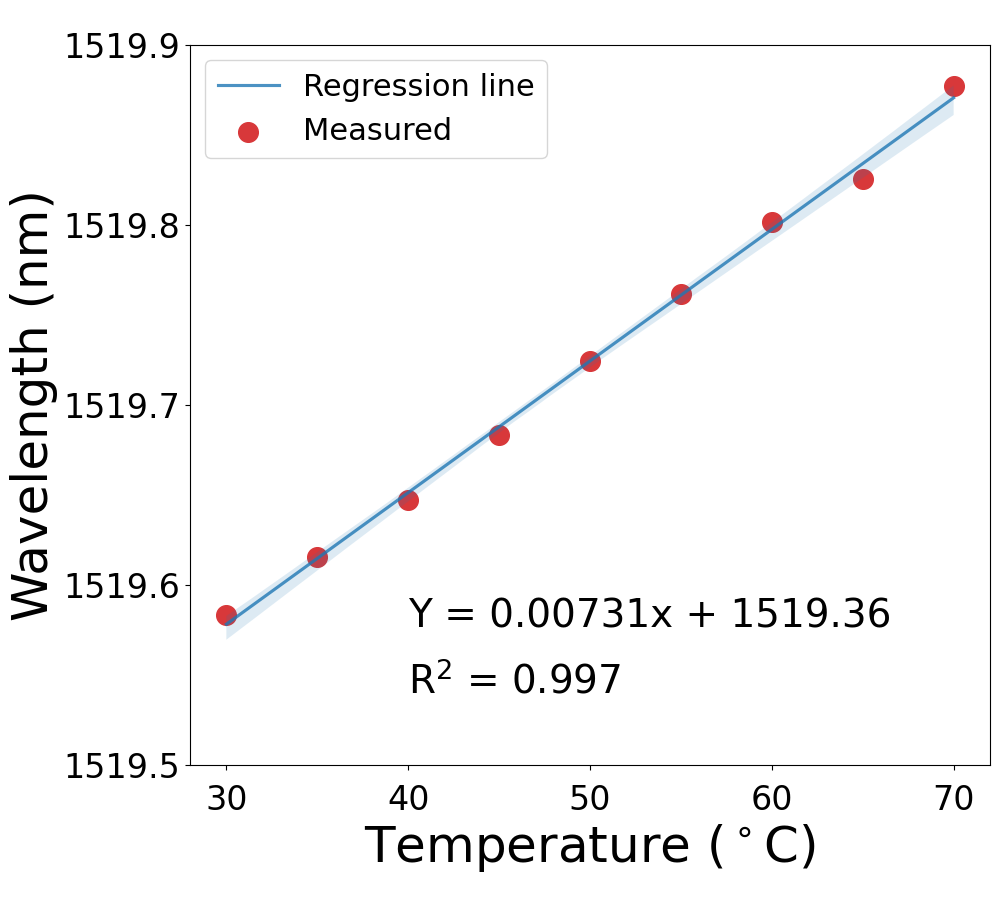}
			\label{fig.12cm.fitting_t}
	\end{minipage}}
	\subfigure[16 cm]{
		\begin{minipage}[t]{0.32\columnwidth}
			\flushleft
			\includegraphics[width=1\linewidth]{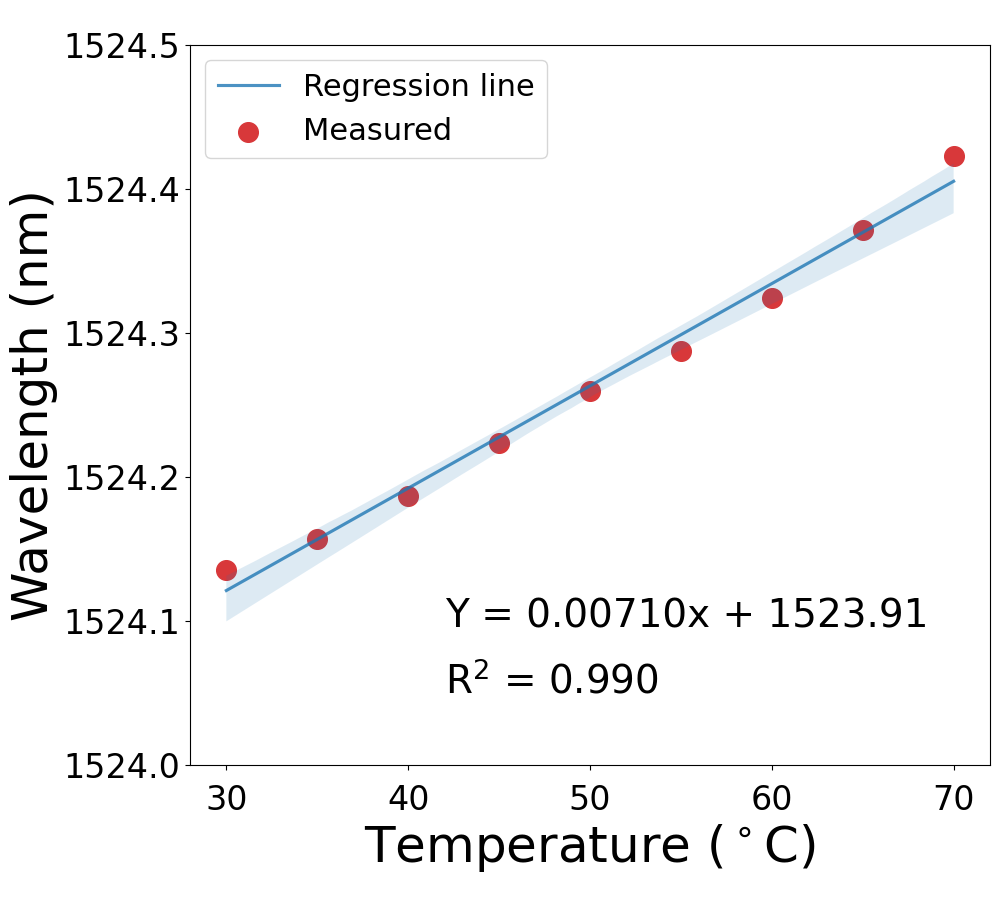}
			\label{fig.16cm.fitting_t}
	\end{minipage}}
	\caption{Temperature measurement results for the MMF with different lengths. Measured spectral dependence on temperature: (a) 2 cm, (b) 4 cm, (c) 8 cm, (d) 12 cm, and (e) 16 cm. Spectral dip shifts as a function of temperature: (f) 2 cm, (g) 4 cm, (h) 8 cm, (i) 12 cm, and (j) 16 cm.}
	\label{fig.lengths.temp}
\end{figure}

\begin{figure}[]
	\vspace{-0.35cm}
	\subfigtopskip=2pt
	\subfigbottomskip=2pt
	\subfigcapskip=-5pt
	\flushright
	\subfigure[4 cm]{
		\begin{minipage}[t]{0.32\columnwidth}
			\includegraphics[width=1\linewidth]{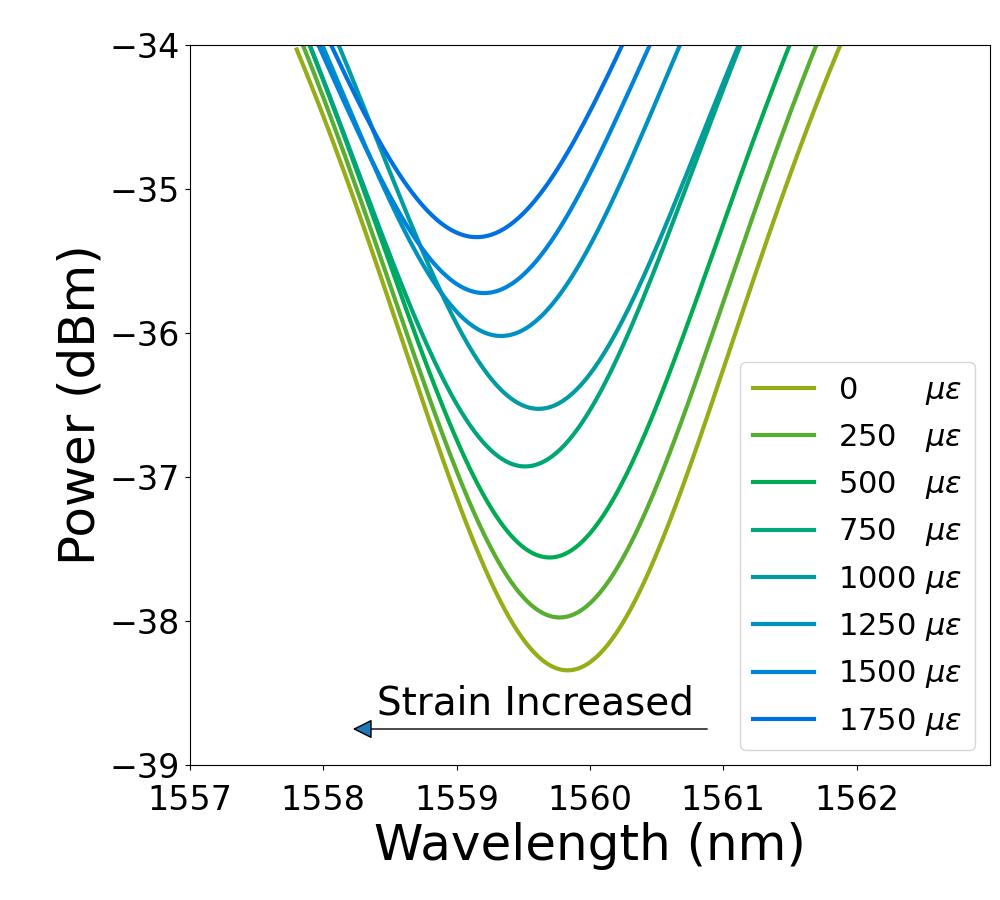}
			\label{fig.4cm.wavelength_strain}
	\end{minipage}}
	\subfigure[8 cm]{
	\begin{minipage}[t]{0.32\columnwidth}
		\includegraphics[width=1\linewidth]{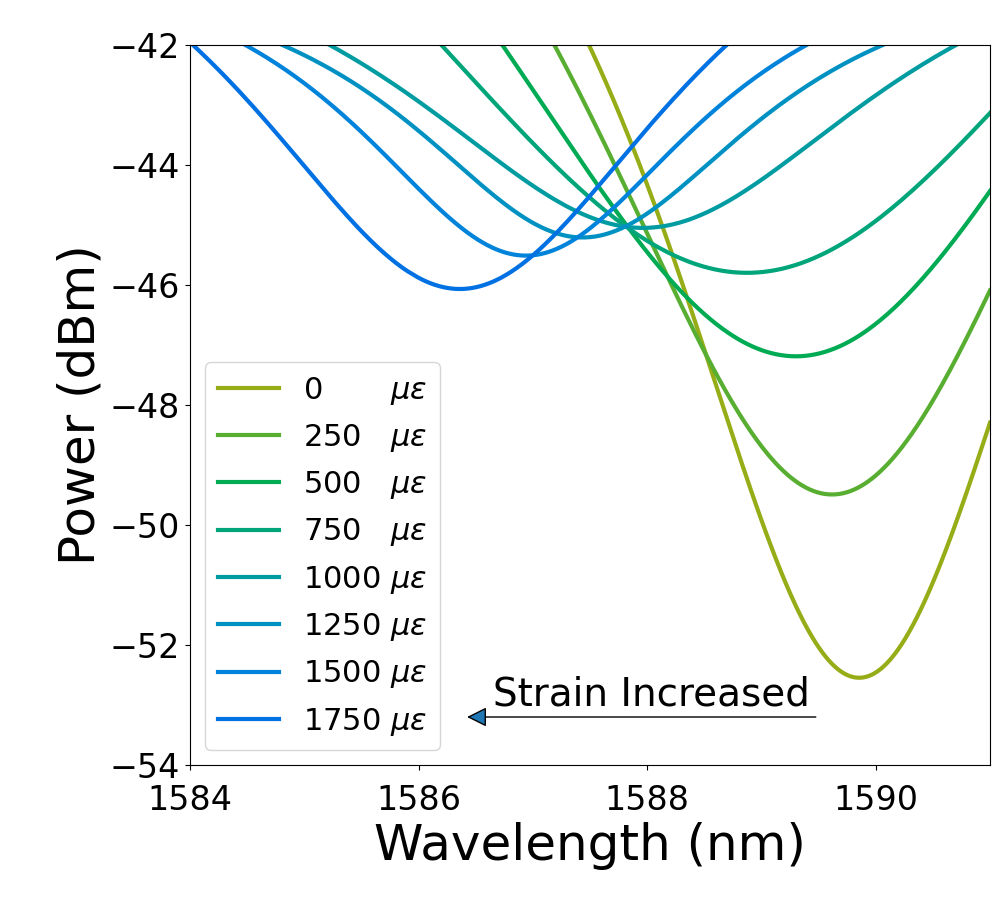}
		\label{fig.8cm.wavelength_strain}
	\end{minipage}}
	\subfigure[12 cm]{
	\begin{minipage}[t]{0.32\columnwidth}
		\includegraphics[width=1\linewidth]{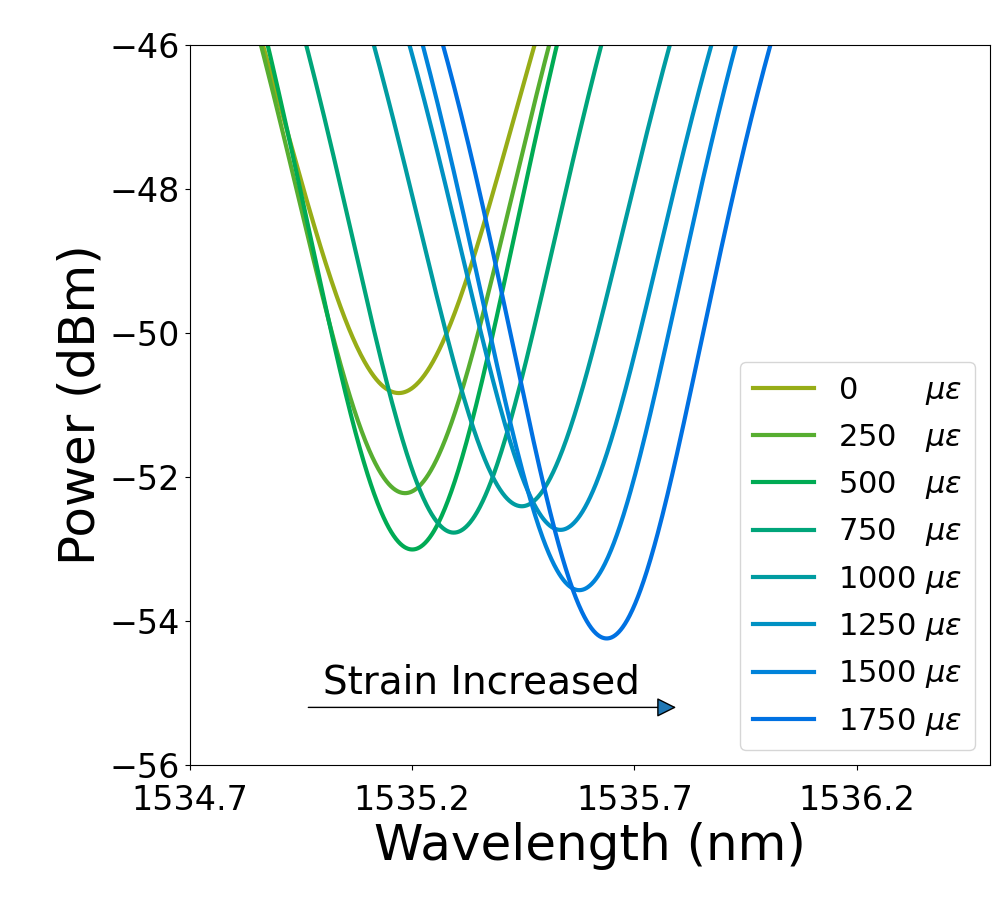}
		\label{fig.12cm.wavelength_strain}
	\end{minipage}}
\\
	\subfigure[4 cm]{
		\begin{minipage}[t]{0.32\columnwidth}
			\includegraphics[width=1\linewidth]{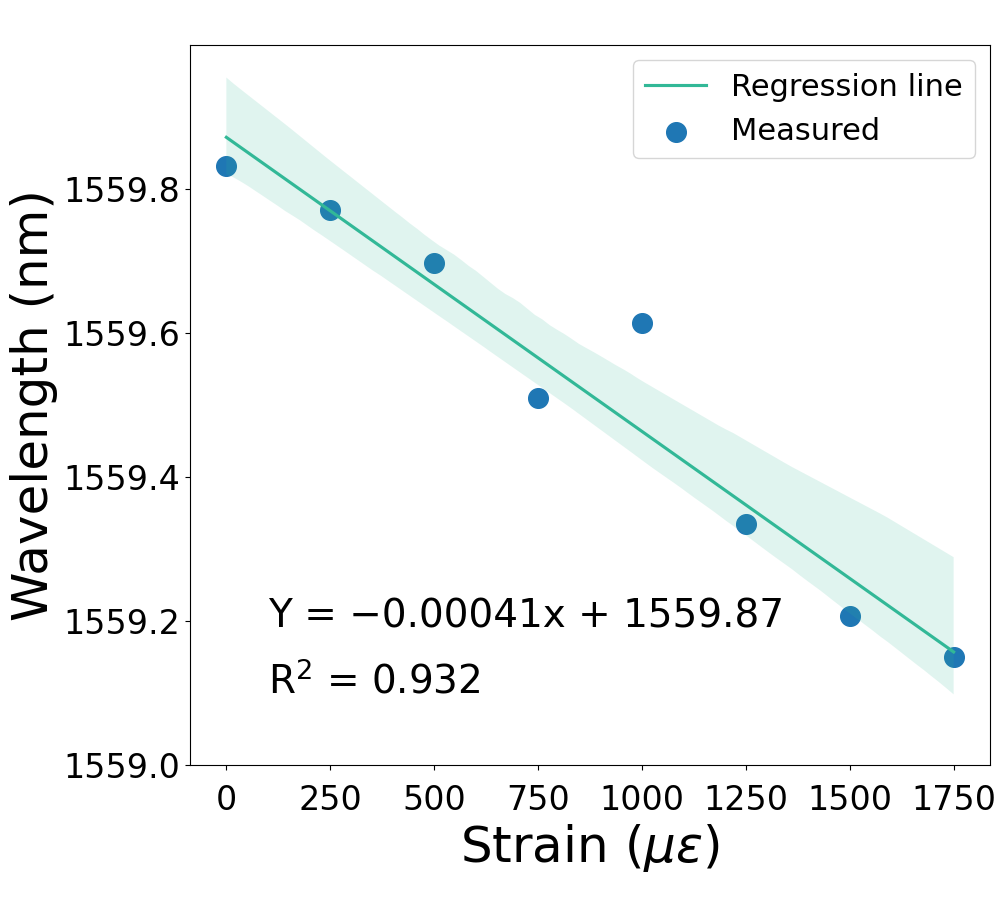}
			\label{fig.4cm.fitting_strain}
	\end{minipage}}
	\subfigure[8 cm]{
		\begin{minipage}[t]{0.32\columnwidth}
			\includegraphics[width=1\linewidth]{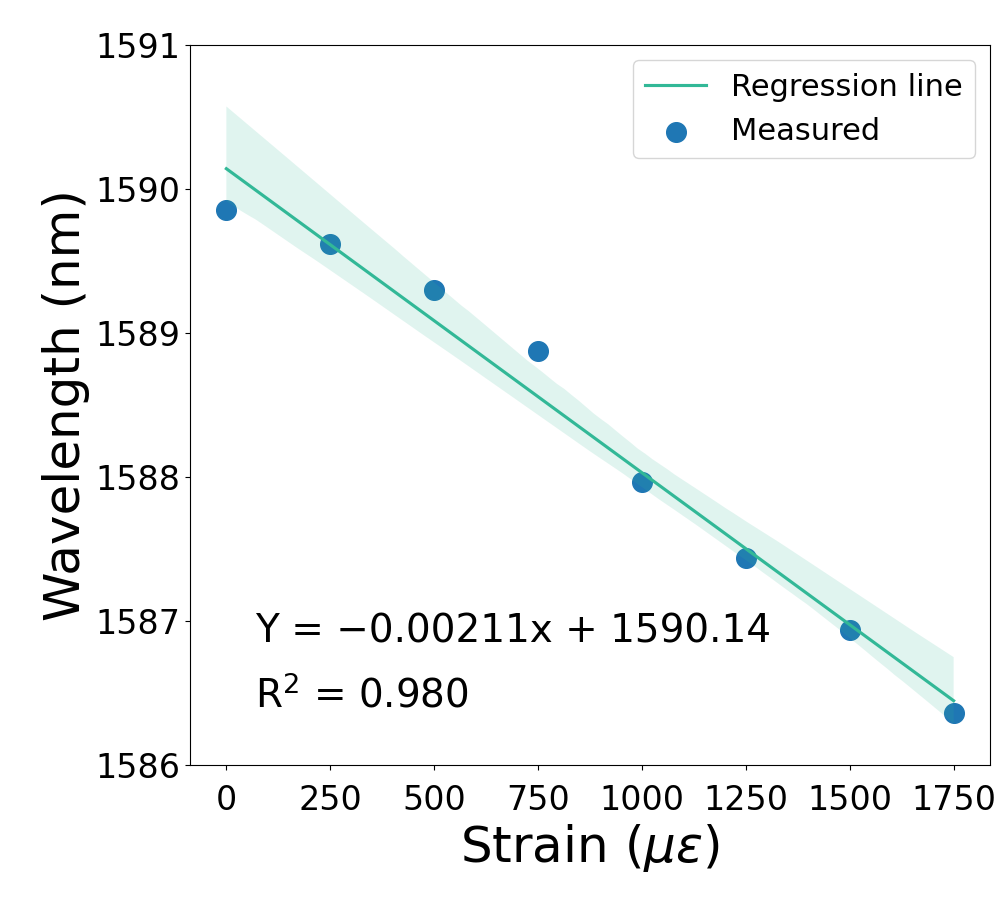}
			\label{fig.8cm.fitting_strain}
	\end{minipage}}
	\subfigure[12 cm]{
		\begin{minipage}[t]{0.32\columnwidth}
			\includegraphics[width=1\linewidth]{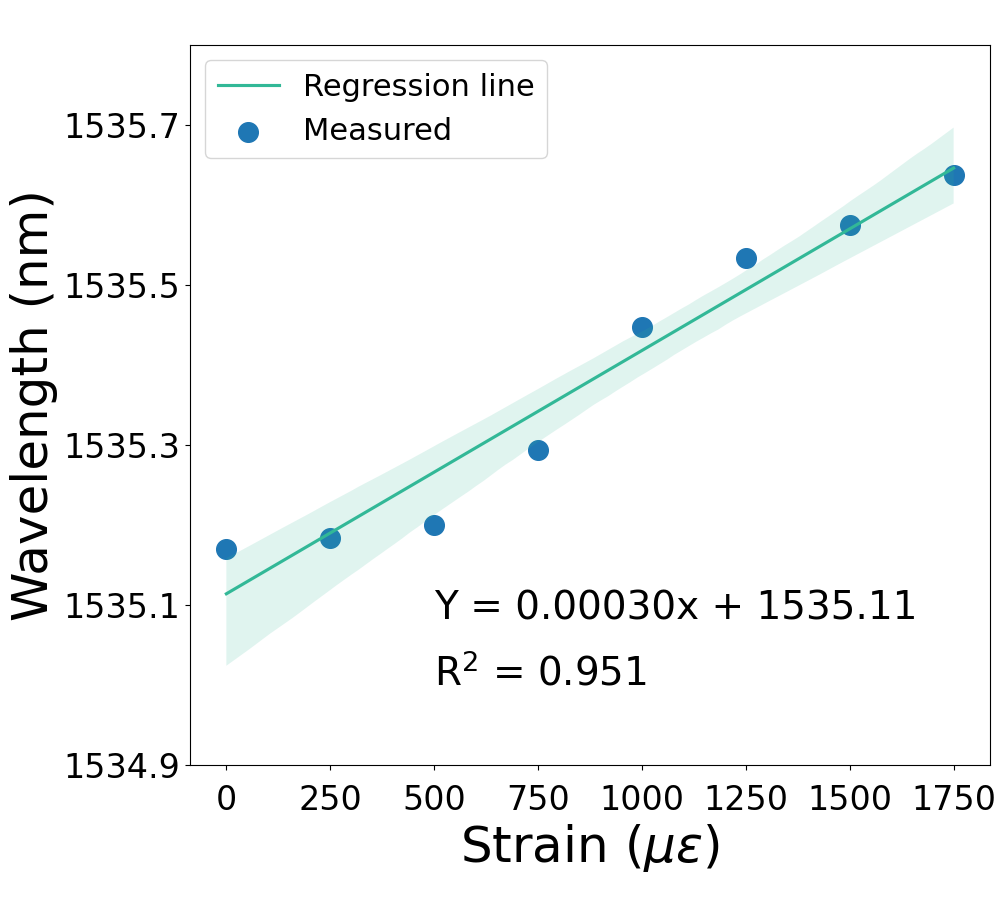}
			\label{fig.12cm.fitting_strain}
	\end{minipage}}
	\caption{Strain measurement results for the MMF with different lengths. (a) 4 cm, (b) 8 cm, and (c) 12 cm. Dip wavelength versus strain: (d) 4 cm, (e) 8 cm, and (f) 12 cm.}
	\label{fig.lengths.strain}
\end{figure}

Figure~\ref{fig.lengths.temp} shows the temperature sensing results. The spectral dip exhibits a redshift in the wavelength domain when the temperature increases for 4-cm-long, 8-cm-long, and 12-cm-long MMFs, as shown in Fig.~\ref{fig.4cm.wavelength_t}, ~\ref{fig.8cm.wavelength_t}, and ~\ref{fig.12cm.wavelength_t}, and the resulting wavelength shift of the dip against the temperature variation is plotted and fitted in Fig.~\ref{fig.4cm.fitting_t}, ~\ref{fig.8cm.fitting_t}, ~\ref{fig.12cm.fitting_t}. The fitting results indicate that the relevant temperature sensitivities are determined to be 8.20~pm/$^{\circ}$C, 7.98~pm/$^{\circ}$C, and 7.31~pm/$^{\circ}$C.
The strain sensing results are presented in Fig.~\ref{fig.lengths.strain}. The spectral dip has a blueshift with the strain applying for the 4-cm-long and 8-cm-long MMF but has a redshift for the 12-cm-long MMF, as shown in Fig.~\ref{fig.4cm.wavelength_strain}, ~\ref{fig.8cm.wavelength_strain}, ~\ref{fig.12cm.wavelength_strain}.
The related resulting wavelength shift of the dip against the increasing strain is plotted and fitted in Fig.~\ref{fig.4cm.fitting_strain}, ~\ref{fig.8cm.fitting_strain}, ~\ref{fig.12cm.fitting_strain}, and the measured strain sensitivities are -0.41~pm/$\upmu\upvarepsilon$, -2.11~pm/$\upmu\upvarepsilon$, and 0.30~pm/$\upmu\upvarepsilon$.

The obtained temperature and strain dependences on the lengths of MMFs are summarized in Table~\ref{table.length}. The temperature and strain dependences are both verified with 4~cm, 8~cm, and 12~cm. The results show that the strain sensitivity is independent of the MMF length, but the temperature sensitivity decreases with increasing MMF length. Based on the results, two more MMF lengths are studied to further enhance the results. Figures~\ref{fig.2cm.wavelength_t} and ~\ref{fig.16cm.wavelength_t} show that the spectral dip exhibits a redshift as well for 2-cm-long and 16-cm-long MMF, respectively. The obtained temperature sensitivities are 9.17~pm/$^{\circ}$C and 7.10~pm/$^{\circ}$C, as shown in Fig.~\ref{fig.2cm.fitting_t} and ~\ref{fig.16cm.fitting_t}, which strengthens the conclusion that the increasing length of MMF results in lower temperature sensitivity.  

\begin{table}[]
	\renewcommand\arraystretch{1.25}
	\centering
	\caption{Temperature and strain sensitivities measured for different lengths of the MMF.}
	\begin{tabular}{ccccc}
		\toprule
		Fiber Type & Dependence & Length & Sensitivity & $R^2$ \\
		\midrule
		\multirow{8}{*}{\makecell[l]{ \\ \\NA = 0.22 \\Length = 8 cm}} & \multirow{3}{*}{\makecell[c]{ \\ \\ \\Temperature}} & 2 cm & 9.17 pm/$^{\circ}$C & 0.986 \\
		& & 4 cm & 8.20 pm/$^{\circ}$C & 0.994\\
		& & 8 cm & 7.98 pm/$^{\circ}$C & 0.990\\
		& & 12 cm & 7.31 pm/$^{\circ}$C & 0.997\\
		& & 16 cm & 7.10 pm/$^{\circ}$C & 0.990\\
		\cmidrule{2-5}
		& \multirow{3}{*}{\makecell[c]{ \\Strain}} & 4 cm & -0.41 pm/$\upmu\upvarepsilon$ & 0.932\\
		& & 8 cm & -2.11 pm/$\upmu\upvarepsilon$ & 0.980\\
		& & 12 cm & 0.30 pm/$\upmu\upvarepsilon$ & 0.951\\
		\bottomrule
	\end{tabular}	
	\label{table.length}
\end{table}

\subsection{Strain-insensitive Temperature Measurement}
Founded on the comprehensive study of temperature and strain sensors based on the SMS structure using standard MMFs, it implies that the strain sensitivity is relatively low. Therefore, a strain-insensitive temperature sensor using standard MMF is proposed in this work. 

\begin{figure}[h]
	\centering
	\vspace{-0.35cm}
	\subfigtopskip=2pt
	\subfigbottomskip=2pt
	\subfigcapskip=-5pt
	\subfigure[]{
		\begin{minipage}[t]{0.4\columnwidth}
			\centering
			\includegraphics[width=1\linewidth]{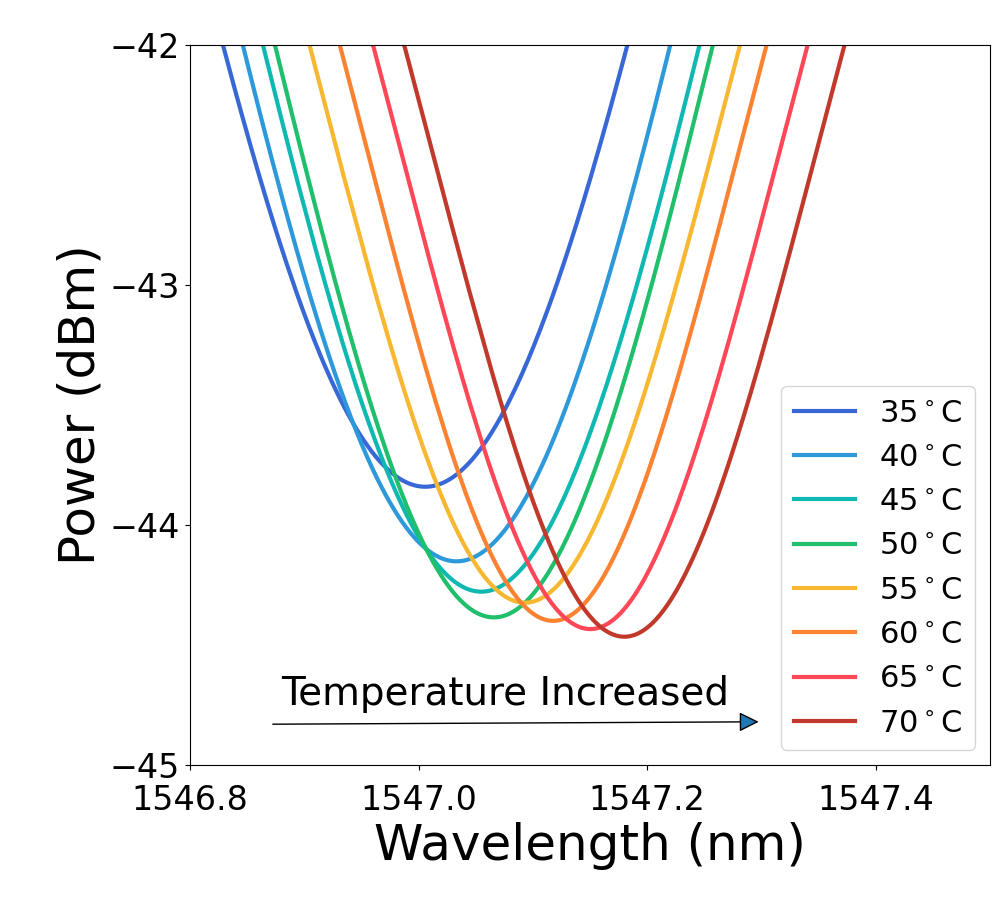}
			\label{fig.insensitive.0}
	\end{minipage}}\quad
	\subfigure[]{
		\begin{minipage}[t]{0.4\columnwidth}
	    	\centering
			\includegraphics[width=1\linewidth]{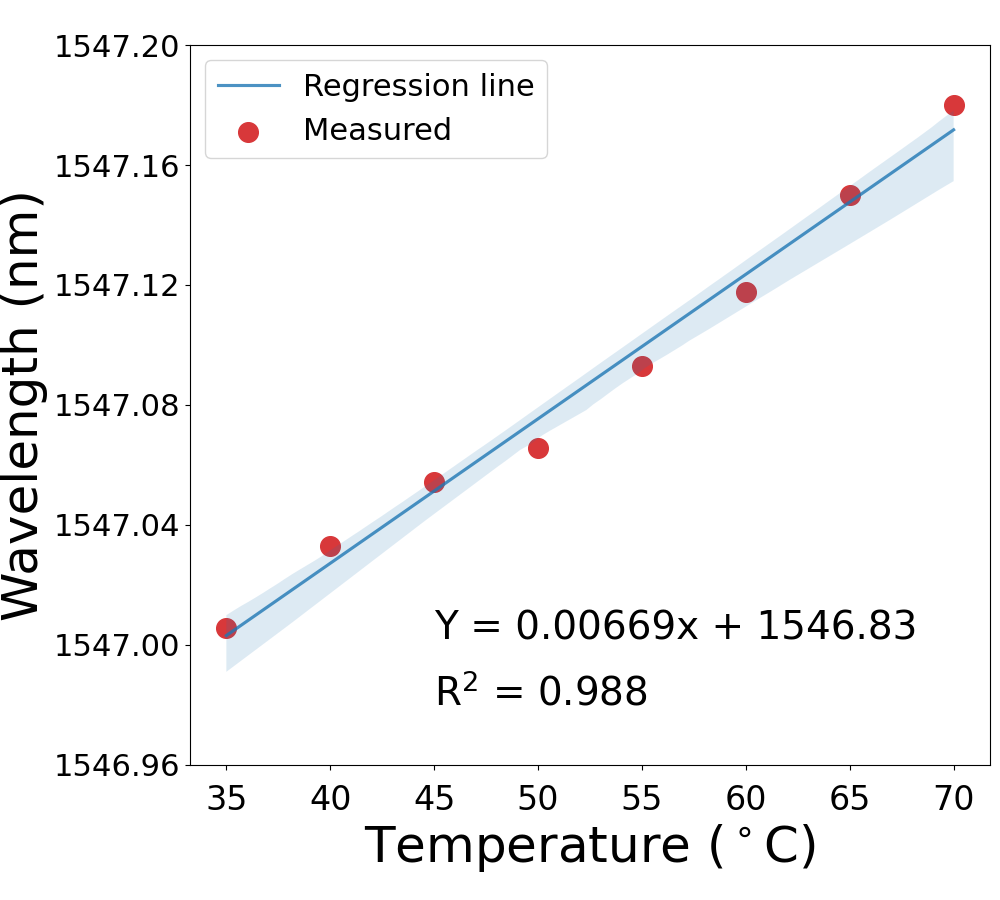}
			\label{fig.insensitive.1}
	\end{minipage}}
	\\
	\subfigure[]{
		\begin{minipage}[t]{1\columnwidth}
			\centering
			\includegraphics[scale=0.3]{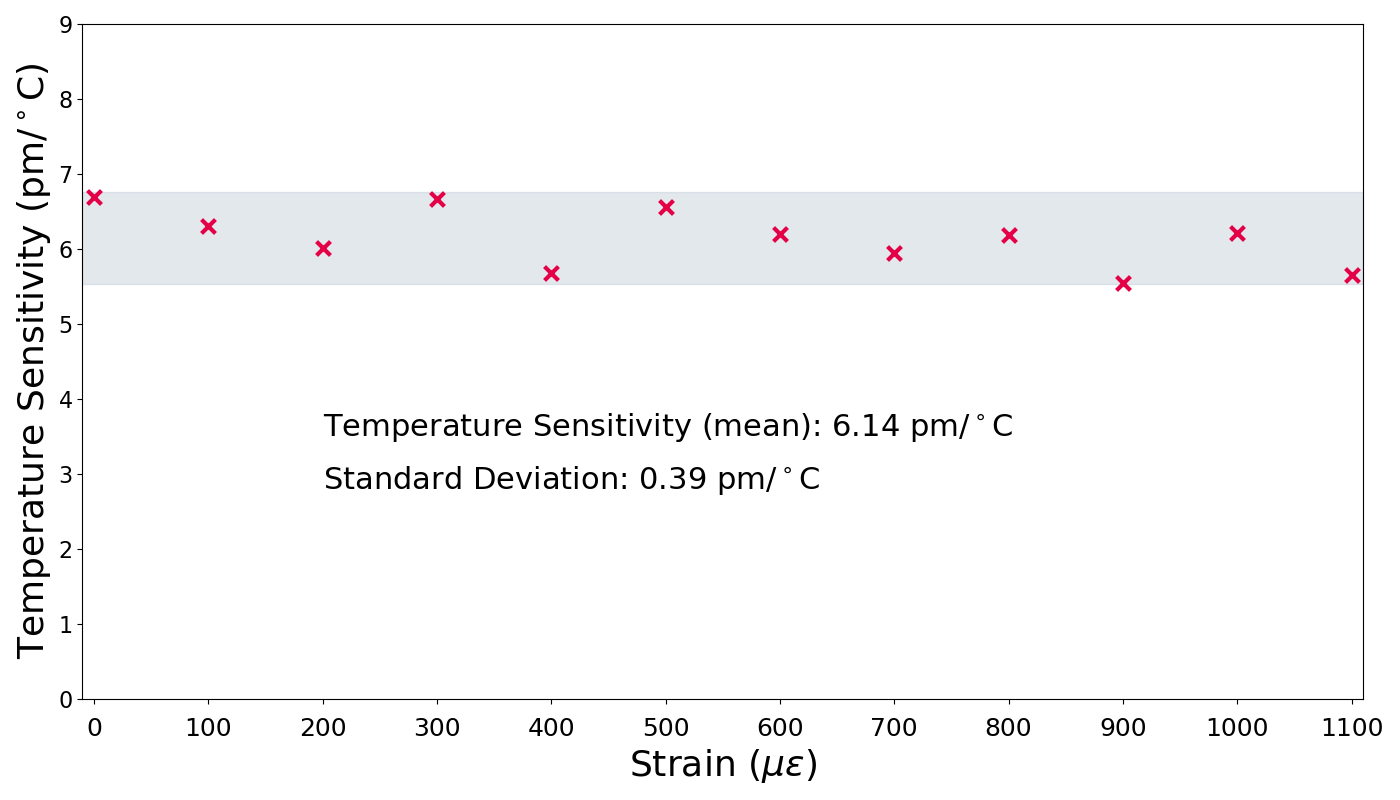}
			\label{fig.insensitive.all}
	\end{minipage}}
	\caption{(a) Measured spectral dependence on temperature when no strain is applied, (b) spectral dip shifts as a function of temperature when no strain is applied, (c) measured temperature sensitivity plotted as a function of strain. The gray area indicates the $\pm$10$\%$ range of the mean temperature sensitivity. 
	}
	\label{fig.insensitive}
\end{figure}

The measurement is performed with the same setup as shown in Fig.~\ref{fig:setup}. According to the configuration of the setup, a 25-cm-long MMF section is placed on the heating plate while the axial strain is applied, which is in the range of 0 to 1100~$\upmu\upvarepsilon$  with steps of 100~$\upmu\upvarepsilon$. 
First, no strain is applied, the temperature measurement is carried out in the same temperature range of 35 to 70$^{\circ}$C with steps of 5$^{\circ}$C. Then, a temperature measurement is completed under the same condition for each strain step. 
Figure~\ref{fig.insensitive.0} shows the measured dependence of the spectral dip on temperature when no strain is applied. The measured temperature sensitivity is 6.69~pm/$^{\circ}$C with an $R^2$ value of 0.988, as shown in Fig.~\ref{fig.insensitive.1}. Compared to Table~\ref{table.length}, as the same MMF with 25~cm is used, it further enhances the conclusion that the longer MMF section leads to lower temperature sensitivity.

The measured results are summarized in Table~\ref{table.insensitive}, and they show that each temperature measurement exhibits high linearity.
As plotted in Fig.~\ref{fig.insensitive.all}, the temperature sensitivities are distributed within $\pm$10$\%$ of the mean temperature sensitivity while the strain is applied in the range from 0 to 1100~$\upmu\upvarepsilon$, which indicates that the temperature sensitivity exhibits good stability and insensitivity against strain. 
The average measured temperature sensitivity is 6.14~pm/$^{\circ}$C with a standard deviation of 0.39~pm/$^{\circ}$C.

\begin{table}[h]
	\renewcommand\arraystretch{1.25}
	\centering
	\caption{Temperature sensitivity measured while strain is applied}
	\begin{tabular}{ccc}
		\toprule
		Applied Strain ($\upmu\upvarepsilon$) & Temperature Sensitivity (pm/$^{\circ}$C) & $R^2$ \\
		\midrule
		0 & 6.69 & 0.988\\
		100 & 6.31 & 0.998\\
		200 & 6.01 & 0.996\\
		300 & 6.67 & 0.998\\
		400 & 5.68 & 0.993\\
		500 & 6.56 & 0.997\\
		600 & 6.19 & 0.998\\
		700 & 5.94 & 0.993\\
		800 & 6.18 & 0.991\\
		900 & 5.54 & 0.997\\
		1000 & 6.21 & 0.999\\
		1100 & 5.65 & 0.988\\
		\bottomrule
	\end{tabular}	
	\label{table.insensitive}
\end{table}

\section{Conclusion}
A strain-insensitive temperature sensor was proposed and investigated experimentally based on comprehensive research of three characteristics of the standard MMF. In this work, different commonly used standard MMFs were studied for temperature and strain sensing. The characterization study of these MMFs concludes:
\begin{enumerate}
	\item The larger core diameter of the MMF leads to higher temperature sensitivity but lower strain sensitivity (absolute values).
	\item The measured temperature sensitivity is independent of the NA of the MMF section, while the higher NA of the MMF leads to higher absolute value of strain sensitivity.
	\item The longer MMF section leads to lower temperature sensitivity, while the measured strain sensitivity is independent of the length of the MMF section.
\end{enumerate}
The strain-insensitive temperature sensor was tested based on these results. The applied strain range is from 0 to 1100~$\upmu\upvarepsilon$ with steps of 100~$\upmu\upvarepsilon$, and the mean measured temperature sensitivity is 6.14~pm/$^{\circ}$C with a standard deviation of 0.39~pm/$^{\circ}$C. We believe that our analysis may contribute to the fundamental discussion of whether the temperature and strain sensitivities of the SMS fiber sensor are dependent on the length of the MMF based on the theoretical analysis in ~\cite{kumar2003}. Thus, this work will be a valuable and important guideline for developing MMI-based fiber sensors in the near future.

\bibliographystyle{unsrt}
\bibliography{Manuscript_arxiv}
\end{document}